\documentclass[aps,twocolumn,noshowpacs, showkeys, pra, superscriptaddress, floatfix, nobalancelastpage, notitlepage, 10pt]{revtex4-1}

\usepackage{graphicx}% Include figure files
\usepackage{dcolumn}% Align table columns on decimal point
\usepackage{bm}% bold math
\usepackage{bbm}
\usepackage{quantikz}
\usepackage{amsmath,amssymb} 
\usepackage{amsthm,epsfig,tikz}
\usepackage[justification=justified]{caption}
\usepackage{xcolor}
\usepackage{verbatim}
\usepackage[normalem]{ulem}
\usepackage[rightcaption]{sidecap}
\usepackage{subcaption}
\usepackage{svg}
\usepackage{booktabs} 
\usepackage{forest,adjustbox}
\usepackage{multirow}
\usepackage{appendix} 
\usepackage{hyperref}  
\hypersetup{colorlinks,citecolor=blue,urlcolor=blue,linkcolor =blue,hypertexnames=true,pdfproducer={Me}}
\usepackage{tikz}
\usepackage{amsfonts}
\usepackage{booktabs}
\usepackage{siunitx}
\usepackage{float}
\usepackage{ragged2e} 
\makeatletter
\let\newfloat\newfloat@ltx
\makeatother
\usepackage{algorithm}
\usepackage{algpseudocode}

\algrenewcommand{\algorithmiccomment}[1]{\hskip2pt\textbackslash\textbackslash \ #1}

\def\bra#1{\mathinner{\langle{#1}|}}
\def\ket#1{\mathinner{|{#1}\rangle}}

\renewcommand{\part}[2]{\frac{\partial #1}{\partial #2}}

\begin{document}

\preprint{JCIM ACS/123-QED} % changed to ACS as discussed in ourt last call

\title{Robust Quantum Reservoir Computing for Molecular Property Prediction}
%\author{Authors}

\author{Daniel Beaulieu}
\email{dabeaulieu@deloitte.com}
\affiliation{Deloitte Consulting LLP}

\author{Milan Kornja\v ca}
\affiliation{QuEra Computing Inc., 1284 Soldiers Field Road, Boston, MA, 02135, USA}

\author{Zoran Krunic}
\affiliation{Amgen, Thousand Oaks, CA 91320 USA}

\author{Michael Stivaktakis}
\affiliation{Technical University of Darmstadt, Darmstadt, Germany}

\author{Thomas Ehmer}
\affiliation{Merck KGaA, Darmstadt, Germany} %https://orcid.org/0000-0002-4586-5361}

\author{Sheng-Tao Wang}
\affiliation{QuEra Computing Inc., 1284 Soldiers Field Road, Boston, MA, 02135, USA}

\author{Anh Pham}
\affiliation{Deloitte Consulting LLP}

\date{\today}

%\captionsetup{justification=raggedright,singlelinecheck=false}
\justifying
%\RaggedRight 

\begin{abstract}
Machine learning has been increasingly utilized in the field of biomedical research to accelerate the drug discovery process. In recent years, the emergence of quantum computing has been followed by extensive exploration of quantum machine learning algorithms. Quantum variational machine learning algorithms are currently the most prevalent but face issues with trainability  due to vanishing gradients. An emerging alternative is the quantum reservoir computing (QRC) approach, in which the quantum algorithm does not require gradient evaluation on quantum hardware. Motivated by the potential advantages of the QRC method, we apply it to predict the biological activity of potential drug molecules based on molecular descriptors. We observe more robust QRC performance as the size of the dataset decreases, compared to standard classical models, a quality of potential interest for pharmaceutical datasets of limited size. In addition, we leverage the uniform manifold approximation and projection technique to analyze structural changes as classical features are transformed through quantum dynamics and find that quantum reservoir embeddings appear to be more interpretable in lower dimensions.
\end{abstract}

\maketitle

%\tableofcontents

\section{\label{sec:intro} Introduction}

Predicting drug activities is an important and difficult task in pharmaceutical discovery. This task has traditionally been performed using a labor-intensive process of repeated trials and errors within a laboratory setting~\cite{liu2016, pharmacokinetics_review}. Recently, machine learning techniques have been employed to accelerate the discovery process~\cite{vamathevan2019, Deng2023, Zeng2022}. Within a machine learning workflow, molecules can be mapped to features using different molecular descriptors, such as physiological, biochemical properties or molecular fingerprints, which can then be used to predict a target variable such as therapeutic efficacy~\cite{rdkit, COMESANA2022123836, CERETOMASSAGUE201558, MELLOR2019121}. However, an immediate challenge is how to extract the relevant molecular features to improve the overall performance of the model, since the structure-property relationship for molecules is complex~\cite{liu2016, vamathevan2019, Vu2023}. Recently, the advent of quantum computing has given rise to a novel machine learning paradigm~\cite{Caro_2022,Schuld_2014,Carleo_2019, Biamonte_2017}. Various quantum machine learning (QML) versions of classical ML models such as generative, support vector machines (SVMs), and neural networks have been explored~\cite{lloyd2020quantumembeddingsmachinelearning,schuld2021supervisedquantummachinelearning,Killoran_2019,huembeli2022entanglementforginggenerativeneural,lloyd2020quantumembeddingsmachinelearning}. Quantum kernel methods, in particular, have shown potential advantage on synthetic datasets of quantum origin ~\cite{Havl_ek_2019,Schuld_2019, Liu_2021, Huang_2021, kornjaca2024}. As a result, there are early attempts to apply QML algorithms for predicting the properties of molecules and pharmacological activity~\cite{Batra2021, Bhatia2023, vakili2024quantumcomputingenhancedalgorithmunveils}.

In this manuscript, we explore the performance of quantum reservoir computing (QRC) for molecular property prediction.  Reservoir computing~\cite{Jaeger2004} addresses the potential drawback of widely employed variational QML models. There, the need to estimate gradients on quantum hardware leads to fundamental trainability issues due to entanglement and noise-induced barren plateaus~\cite{McClean_2018, marrero2021entanglementinducedbarrenplateaus, larocca2024reviewbarrenplateausvariational}. QRC has been proposed as an alternative approach, in which the gradient estimation can be fully off-loaded to classical post-processing~\cite{Fujii_2017, Martinez2021,  Bravo2022,  Senanian2023, kornjaca2024}. The QRC algorithm we employ in this study leverages entangled quantum dynamics of neutral atom arrays, a platform that has already shown promise for quantum simulation, optimization, and machine learning~\cite{Scholl2021, Ebadi2021, Semeghini2021, Ebadi2022, wurtz2023aquila, quera2024, Bluvstein2023, kornjaca2024}. % Subsequently, the embeddings generated from the expectation values of local observables during emulated quantum dynamics are used to train an ensemble of classical models.

We applied the QRC algorithm to the dataset within the Merck Molecular Activity challenge (MMACD)~\cite{merck2012},  with the task of predicting pharmacological activity based on molecular fingerprints~\footnote{The Competition was sponsored in 2012 by Merck \& Co., Inc, located today at 26 E Lincoln Ave, Rahway, NJ 07065, US, USA. Merck KGaA and Merck \& Co.~are separate entities. Merck KGaA is headquartered in Germany, operates globally as EMD in North America, focusing on pharmaceuticals, life sciences, and performance materials, whereas Merck \& Co., headquartered in the USA, operates as MSD outside North America, primarily focusing on pharmaceuticals, vaccines, and animal health.}. We systematically report the performance of the classical and QRC models across different dataset sample sizes and feature choices generated by the quantum reservoir dynamics. We then compare classical features and embedded quantum reservoir features by employing the uniform manifold approximation and projection (UMAP) technique. Our results reveal a slower performance decay with decreasing training dataset size when QRC embeddings are used compared to classical features. The UMAP analysis reveals more structure for QRC-embedded data when projected to low dimensional spaces. Together, the results indicate that the QRC approach has the potential to build a more robust and interpretable model with less training data, which could be relevant for health-related applications.

\begin{figure*}[!ht]
    \centering
    \includegraphics[width=0.9\linewidth]{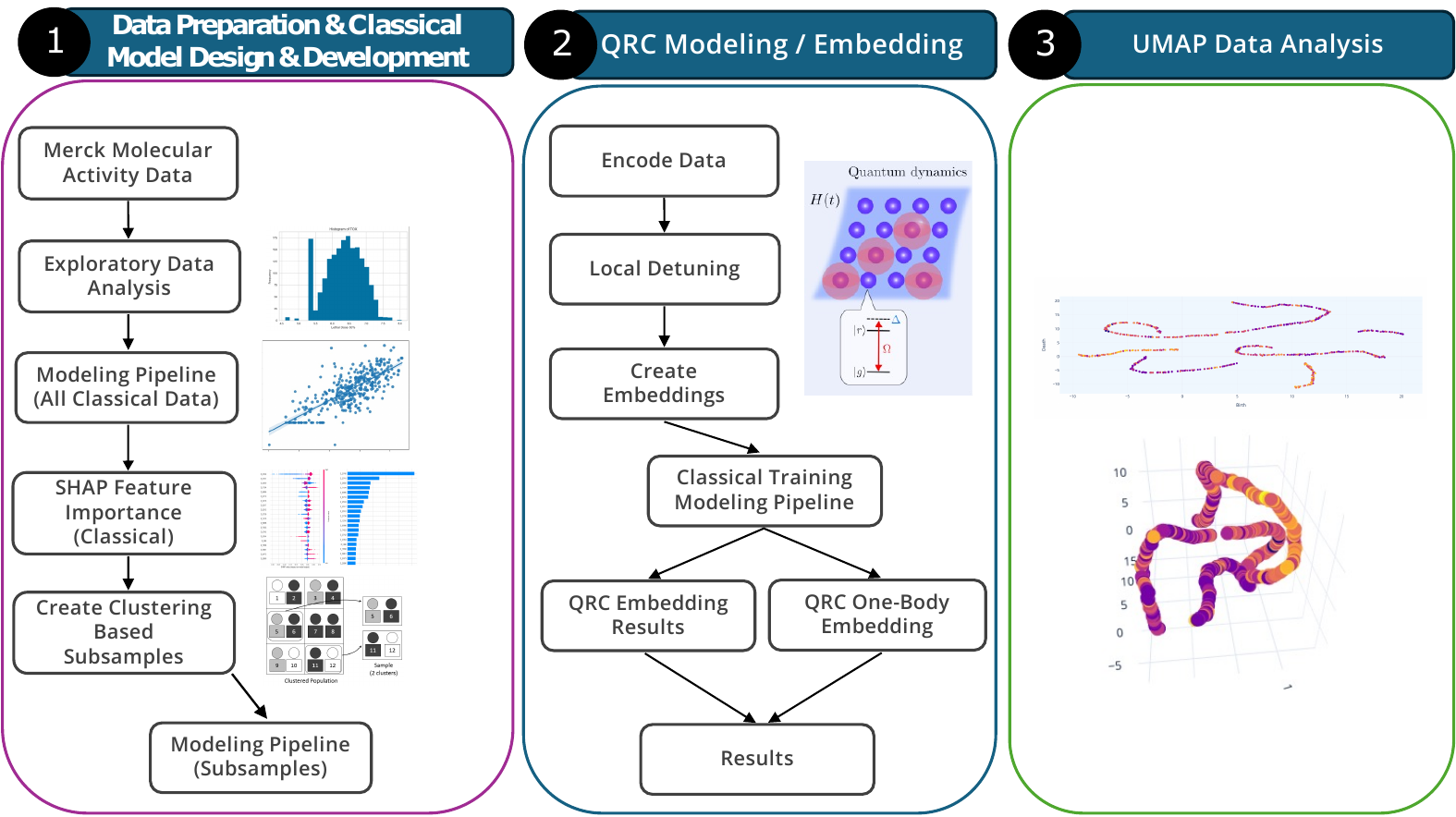}
    \caption{\justifying \textbf{Machine learning modeling workflow and analysis of the different embeddings}. \textbf{(1)} A schematic workflow used to determine the optimal classical model, the feature selection process with the SHAP method to enable QRC simulation, and the subroutine to perform dataset subsampling. \textbf{(2)} An outline of the quantum reservoir computing algorithm. Both two-body observable embeddings and one-body observable embeddings are estimated and utilized to train the classical models. \textbf{(3)} UMAP analysis is used to interpret the difference between classical and quantum features.}
    \label{fig:QRC-workflow}
\end{figure*}

\section{\label{sec:methods} Methods}

\subsection{Molecular dataset}\label{sec:mol_data}
To explore how quantum reservoir computing can be utilized to make predictions of molecular activities, we applied the QRC technique to predict the activity (ACT) values based on molecular descriptors as presented in the MMACD~\cite{merck2012}.  The dataset contains 15 different biological activity prediction problems, each represented by a unique set of molecular descriptors. The MMACD has been used extensively in the machine learning community~\cite{ramsundar2015massivelymultitasknetworksdrug, goh2017chemceptiondeepneuralnetwork, hechtlinger2017generalizationconvolutionalneuralnetworks}. In this manuscript, we perform general modeling and the QRC pipeline to five of the smallest MMACD datasets, 4, 5, 9, 14, and 15, out of the entirety of the MMACD. For our UMAP projection and more detailed analysis, we focus our study on one of the 15 ACT datasets, MMACD 4. MMACD 4 was selected as the ideal candidate for probing the robustness of QRC due to having the smallest sample size. In addition, we chose MMACD 14 to perform our 25 subsample QRC modeling on analysis so as to reduce potential dataset bias for our 25 subsample results.

\subsection{Research Modeling Workflow Overview}
\label{sec:Methods_overview}
As shown in Fig.~\ref{fig:QRC-workflow}, the modeling process starts with data preparation and exploratory data analysis, as described in Appendix \ref{sec:dataprep_appendix}. We then model all available data to select the best candidate model for the raw classical data, as described in Appendix \ref{sec:cand_mods}. The best model is used as a basis for SHAP feature importance selection to select the top 18 features. Clustering based sampling is used to create five subsamples of 100, 200, and 800 records. The modeling pipeline is applied to each such subsampled dataset for several subsample sizes. We feed the subsamples into the Julia QRC simulation pipeline~\cite{kornjaca2024} and create QRC embeddings. The QRC embeddings are fed into the classical modeling pipeline for all subsamples. Subsequent analysis of the embeddings is performed using the UMAP technique~\cite{mcinnes2020umapuniformmanifoldapproximation}.

\subsection{Machine Learning Pipeline} \label{sec:ML_methods}
In this study, we implemented a machine learning pipeline to evaluate the performance of various regression algorithms on the MMACD, as shown in Fig.~\ref{fig:QRC-workflow}. The first step in our pipeline is the pre-processing step that involves inputting missing values using a data standardization procedure described in Appendix \ref{sec:dataprep_appendix}. The next step is to conduct an exploratory data analysis to determine salient features and transform data as necessary, also described in more detail in Appendix \ref{sec:dataprep_appendix}. Subsequently, we bined the pre-processing steps and the specified regression modeling algorithms for fitting the models to the training data, which is described in Appendix \ref{sec:mlpipeline}. The next step in the machine learning pipeline illustrated in Fig.~\ref{fig:QRC-workflow} is the use of the SHAP method~\cite{10.5555/3295222.3295230, shapintro}  to choose the most important features and reduce the feature dimension to 18. We choose 18 data features to balance computational feasibility (of simulating the QRC dynamics) with maintaining the most relevant molecular information. We find that the difference in classical model performance with the top 18 predictors and all predictors is less than 1 percent in modeling MMACD 4 dataset. The biological relevance of our selected features is implicitly addressed through our use of established molecular descriptors from the MMACD dataset.

We then proceed to assign data to clusters of different sizes. These cluster values serve two purposes. First, they ensure a representative subsample of the ensemble from which the data is sampled. Second, they allow us to test the performance of QRC-embedded data versus raw features across various sample sizes. Optionally, we employ the QRC algorithm to create the QRC embeddings, which are used as input training data for our ensemble of classical models. Overall, the goal of this pipeline is to pass the data, either the raw features or QRC embedded features, to a series of regression models, and compare their predictive abilities. In this procedure, we ensure that the data is processed in the same way for a fair comparison at different subsample sizes (100, 200, and 800).

In our pipeline, we evaluated ten different regression algorithms, described in Appendix \ref{sec:mlpipeline}: Random Forest Regressor, AdaBoost Regressor, Gradient Boosting Regressor, Bagging Regressor, SVR, Decision Tree Regressor, Extra Tree Regressor, Linear Regression, SGD Regressor, and KNeighbors Regressor as implemented in the Scikit-learn package~\cite{scikit-learn}. For each algorithm, we compare their predictive ability using the mean squared error (MSE). We iterate over the list of algorithms, fitting each model to the training data, predicting on the test data, and computing the performance metrics. We then decide to report the results for two models -- random forest regressor~\cite{randomforestreg}, as it consistently performed the best across sample sizes and embedding types, and Gaussian radial basis function (RBF) kernel~\cite{gaussianprocessregressor} since it is a representative classical kernel algorithm.

\subsection{Quantum Reservoir Computing}\label{sec:QRC_methods}

The QRC method was applied to the 18 classical features selected from our SHAP analysis, with the numerical simulations of neutral atom dynamics performed with the Bloqade simulator~\cite{bloqade2023quera, QRCtutorials}. The neutral-atom QRC protocol, described in~\cite{kornjaca2024}, encodes the data features into the dynamics of the Rydberg Hamiltonian that governs the analog dynamics of neutral atom arrays~\cite{wurtz2023aquila}:
\begin{align}\label{eq:RydHam}
    H(t)&=\dfrac{\Omega(t)}{2}\sum_j \left(\ket{g_j}\bra{r_j}+\ket{r_j}\bra{g_j}\right)\cr
    &+\sum_{j<k}V_{jk}n_jn_k-\sum_j \left[\Delta_{\mathrm{g}}(t) + f_j\Delta_{\mathrm{l}}(t)\right] n_j.
\end{align}
Here, $\Omega(t)$ represents the global Rabi drive amplitude between a ground ($\ket{g_j}$, $j$ indexes atoms) and an excited state of an atom ($\ket{r_j}$), $n_j=\ket{r_j}\bra{r_j}$, while  $V_{jk}=C/\lVert\mathbf{r}_j-\mathbf{r}_k\rVert^6$ describes the interactions between atoms. The detuning is split into the global term, $\Delta_{\mathrm{g}}(t)$ and the site-dependent term $\Delta_{\mathrm{l}}(t)$, with site modulation $f_j\in[-1,1]$. 

The data features were normalized and encoded into the local detuning pattern, $f_j$, with constant $\Delta_{\mathrm{l}}=6 \,\mathrm{rad}/\mu \mathrm{s}$. We set $\Omega = 2\pi \,\mathrm{rad}/\mu \mathrm{s}$, set the atoms in the linear chain with interatomic distance of $10\, \mathrm{\mu m}$, and probe the quantum dynamics starting from the all-ground state for $4.3 \, \mathrm{\mu s}$, with a step-size of $0.4 \, \mathrm{\mu s}$. The expectation values of the local observables in the computational basis are collected, including one-body $\langle Z_i \rangle$ and two-body $\langle Z_{i}Z_{j} \rangle$ observables. These expectation values form feature embeddings, which are then fed further into the classical model-based postprocessing pipeline. 

\begin{figure*}[!htb]
    \centering
    \begin{subfigure}[b]{0.48\linewidth}
        \centering
        \includegraphics[width=\linewidth]{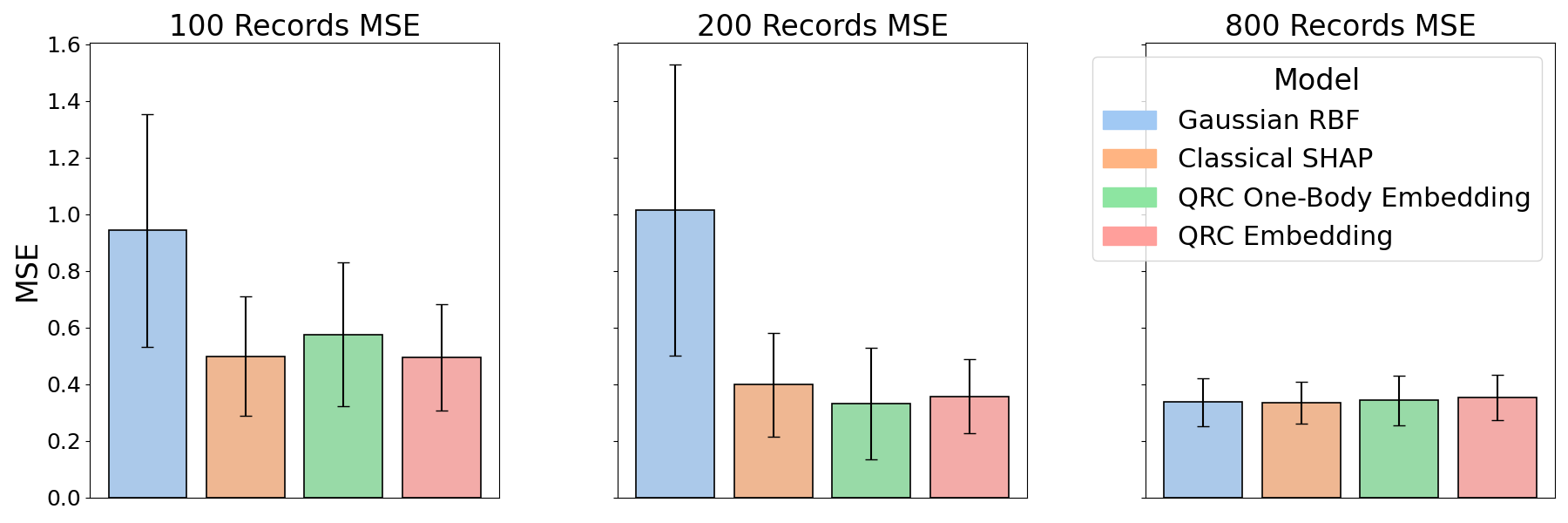}
        \caption{\justifying \textbf{MMACD 4 Results, 5 subsamples}}
        \label{fig:act-4-model-metric-figure}
    \end{subfigure}
    \hfill
    \begin{subfigure}[b]{0.48\linewidth}
        \centering
        \includegraphics[width=\linewidth]{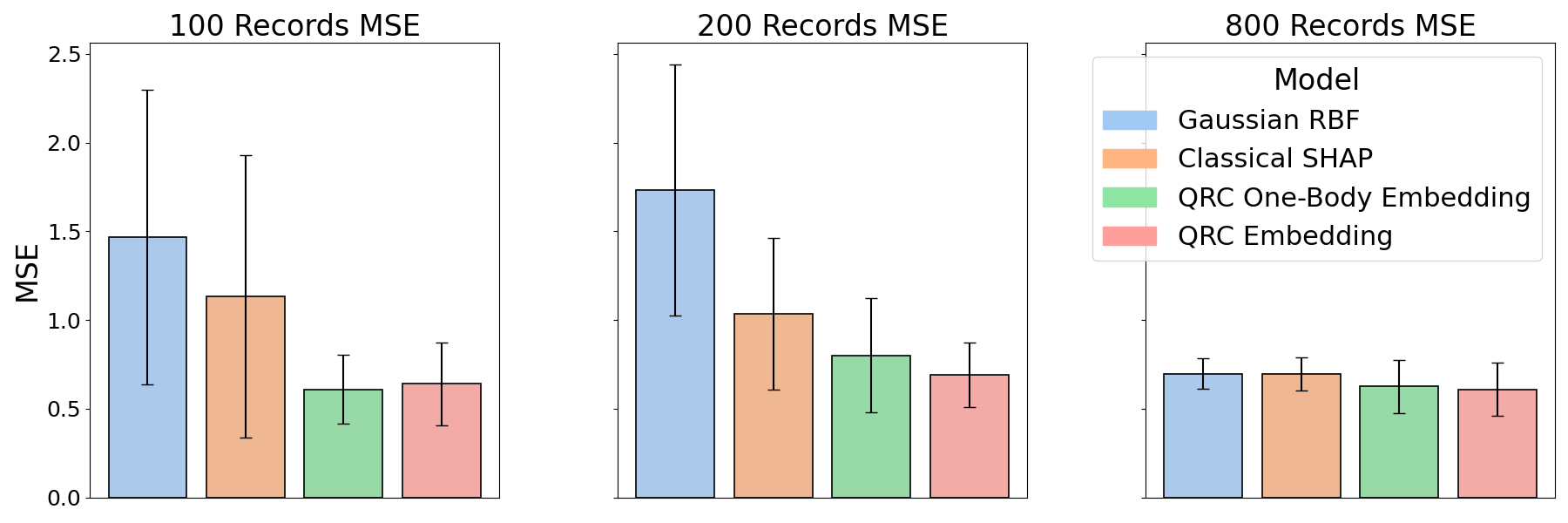}
        \caption{\justifying \textbf{MMACD 5 Results, 5 subsamples}}
        \label{fig:act-5-model-metric-figure}
    \end{subfigure}
    \vspace{0.5cm}
    \begin{subfigure}[b]{0.48\linewidth}
        \centering
        \includegraphics[width=\linewidth]{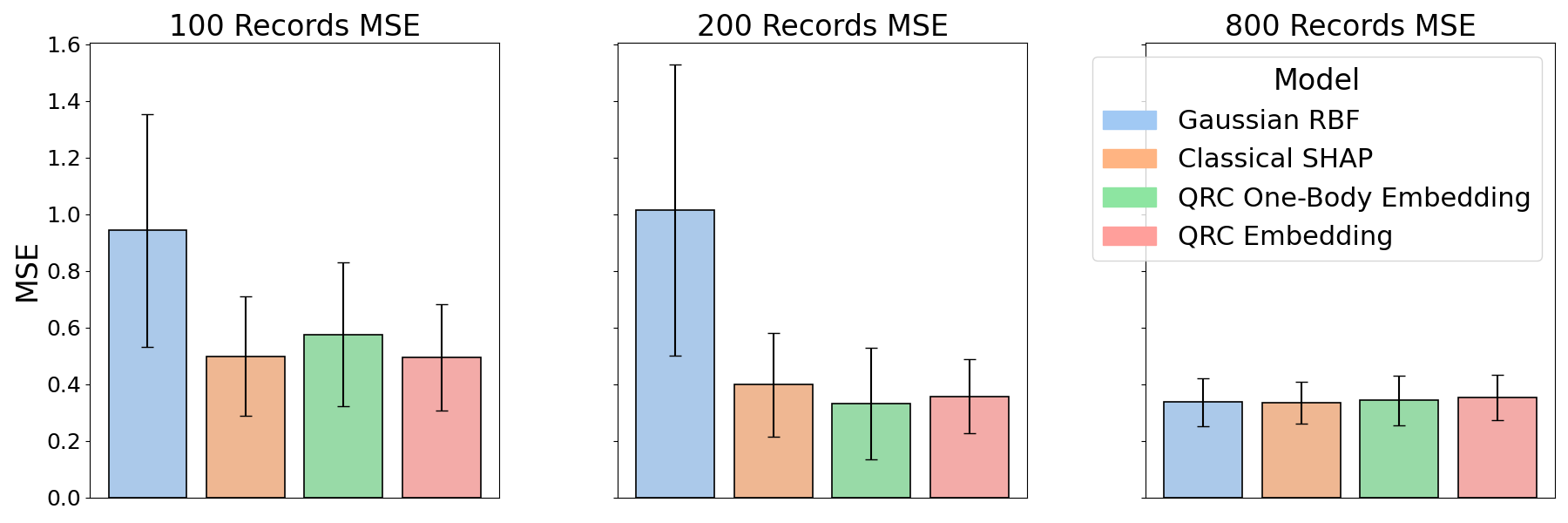}
        \caption{\justifying \textbf{MMACD 9 Results, 5 subsamples}}
        \label{fig:act-9-model-metric-figure}
    \end{subfigure}
    \hfill
    \begin{subfigure}[b]{0.48\linewidth}
        \centering
        \includegraphics[width=\linewidth]{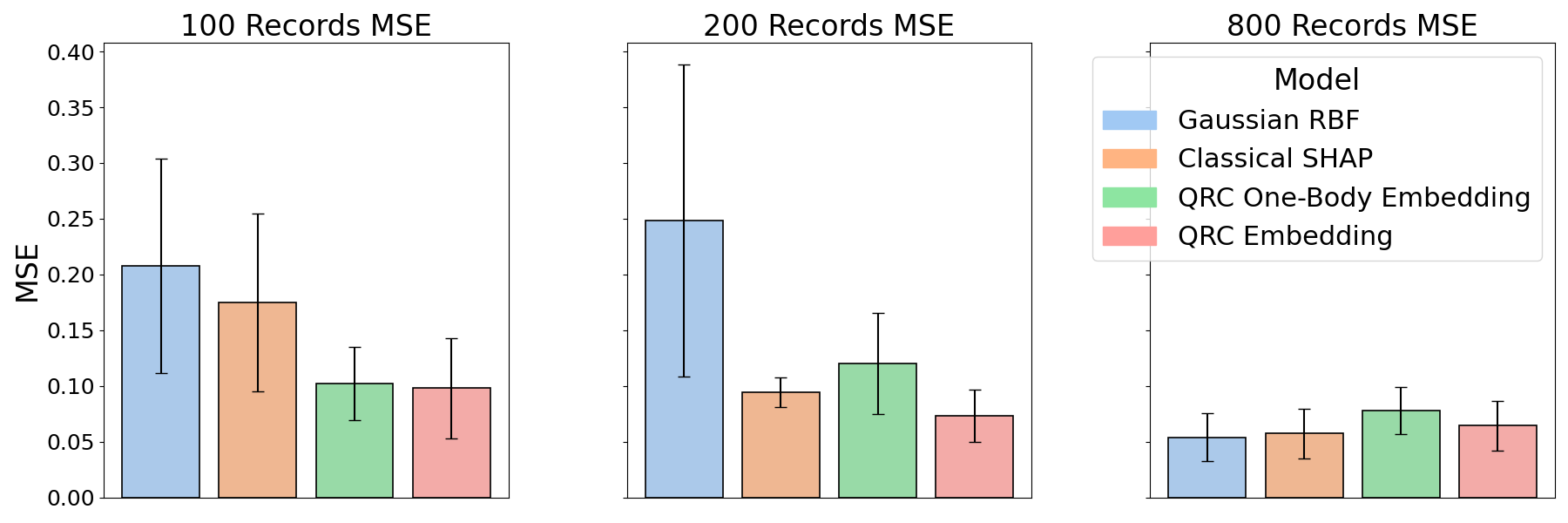}
        \caption{\justifying \textbf{MMACD 14 Results, 5 subsamples}}
        \label{fig:act-14-model-metric-figure}
    \end{subfigure}
    \vspace{0.5cm}
    \begin{subfigure}[b]{0.48\linewidth}
        \centering
        \includegraphics[width=\linewidth]{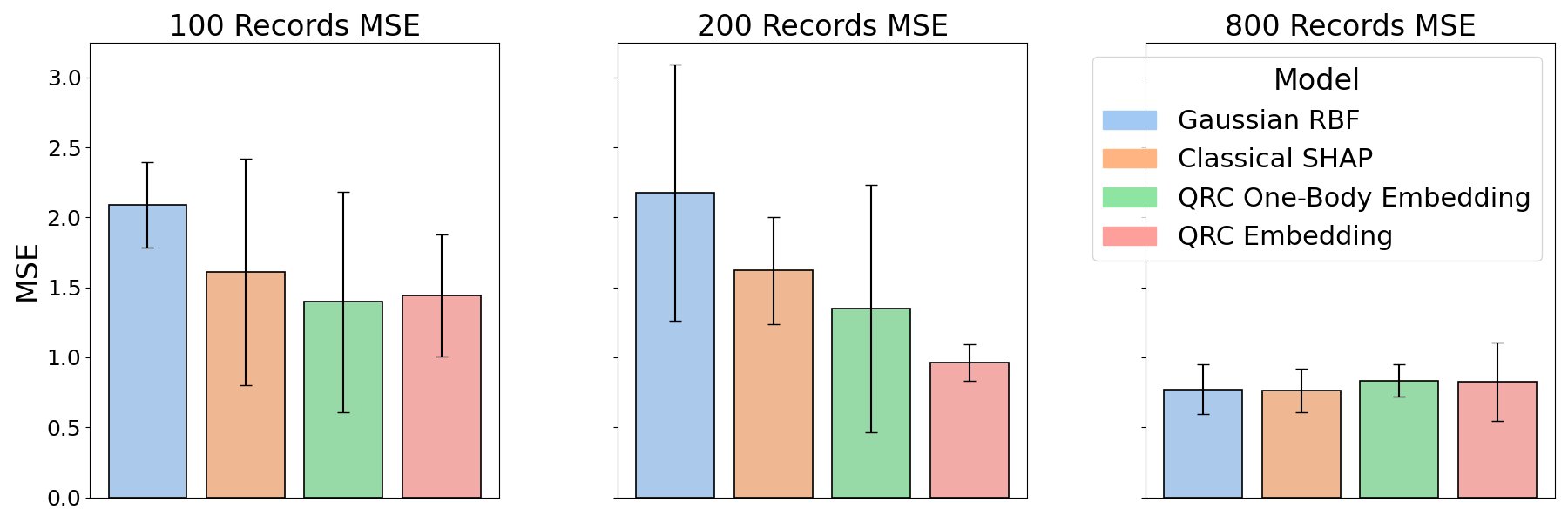}
        \caption{\justifying \textbf{MMACD 15 Results, 5 subsamples}}
        \label{fig:act-15-model-metric-figure}
    \end{subfigure}

    \caption{\justifying \textbf{The performance of the classical and quantum methods on the MMACD datasets.} (a)-(e) Test mean square error (smaller is better) is reported for the five smallest datasets in MMACD and for two classical and two quantum feature methods. In all cases, we report averages across five dataset subsamples, with standard deviation over that sample reported as uncertainty estimate.}
    \label{fig:combined-model-metric-figure}
\end{figure*}

\subsection{Uniform Manifold Approximation and Projection (UMAP) 2D Projection}\label{sec:umapdesc}

To try to understand the embeddings and have better interpretability, we utilized UMAP to project the high dimensional data to low dimensional spaces~\cite{mcinnes2020umapuniformmanifoldapproximation}. The UMAP algorithm is applied to the dataset features to generate 2D embeddings. These embeddings are used to visualize the molecular descriptors and, optionally, to facilitate the training of machine learning models. Using UMAP for dimensionality reduction offers an interpretation of the data's structure in a low-dimensional space, enabling us to identify patterns that are not immediately apparent in high-dimensional space, as demonstrated in previous studies~\cite{Dorrity2020}. Specifically, we projected the data features to 2D with UMAP, to compare the QRC generated feature embeddings with raw features. UMAP projection was guided by several parameters: the number of neighbors was set to 200 to maximize global structure vs local structure in the data; the minimum distance between points in the low-dimensional space was set to 0.9 to control the tightness of UMAP's embedding and maximize global UMAP structure; and the metric for distance calculation chosen is Minkowski, a generalization of both Euclidean distance and Manhattan distance. The same parameters were used to perform the UMAP projection for both the quantum and classical features.

\begin{figure*}[!htb]    
    \centering
    \includegraphics[width=0.27\linewidth]{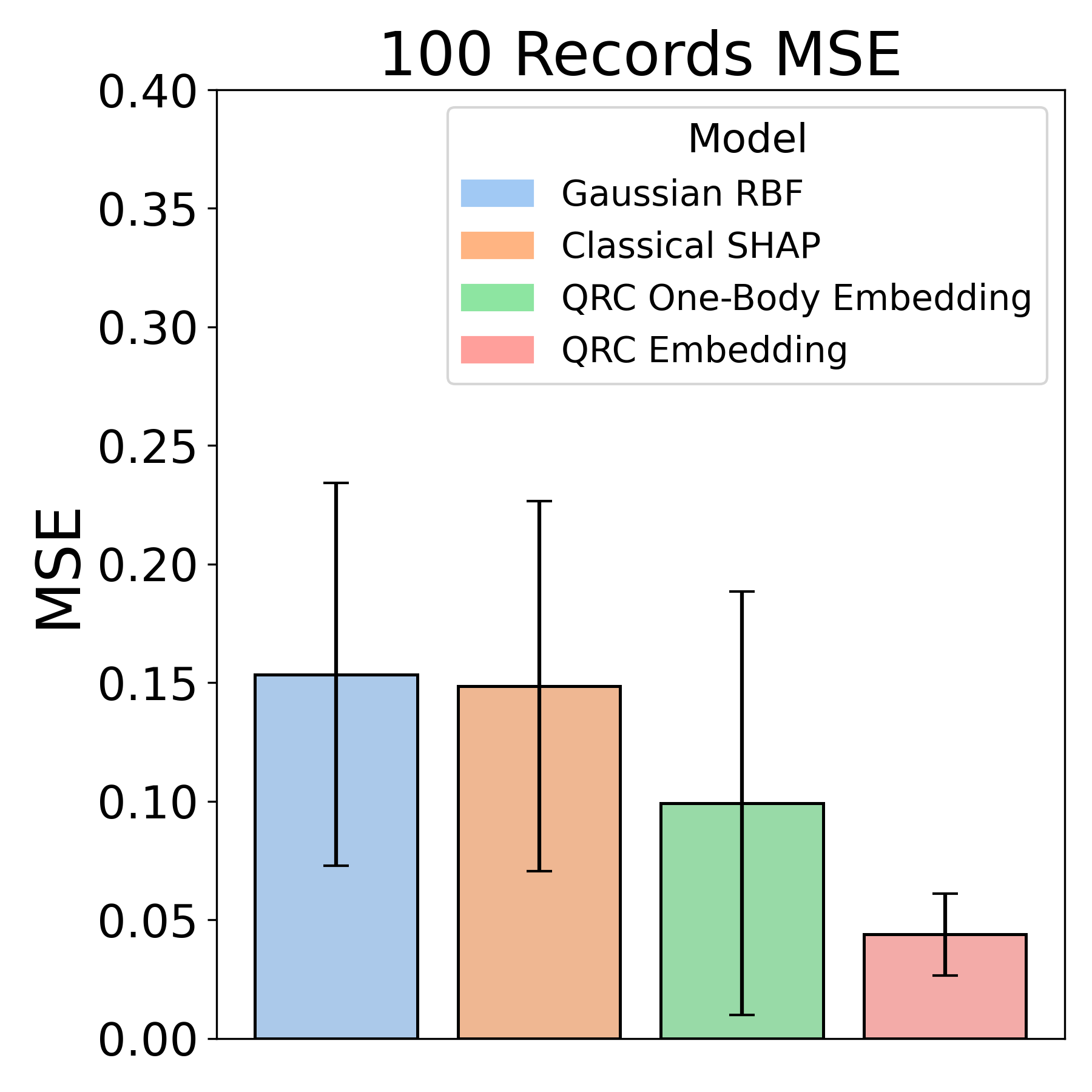}
    \caption{\justifying \textbf{The performance of the classical and quantum methods on MMACD 14 with 25 subsamples of 100 records.} QRC, raw data, and Gaussian RBF results for the MMACD 14 data. Here, we report averages across twenty-five dataset subsamples, with standard deviation over that sample as uncertainty estimate.}
    \label{fig:act14-25subsamp}
\end{figure*}

\begin{figure*}[!htb]    
    \centering
    \includegraphics[width=0.95\linewidth]{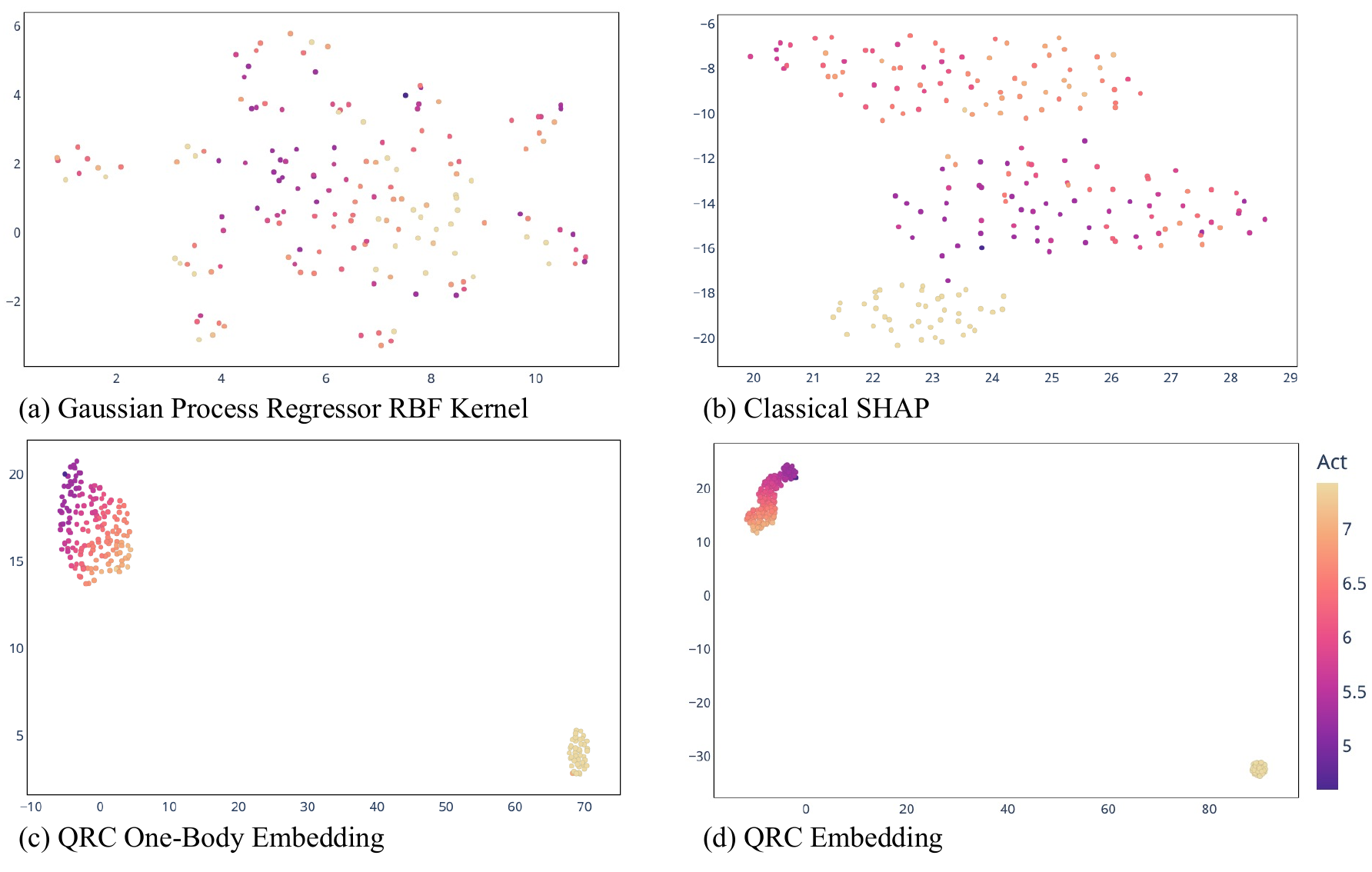}
    \caption{\justifying \textbf{UMAP 2D projections for 200 record samples from MMACD 4 dataset and sub-sample 4.}}
    \label{fig:umap_200recs}
\end{figure*}

\section{Results and Discussion}\label{sec:results_discussion}
\subsection{QRC and classical model performance}

Based on the results of the initial pass of classical data through the modeling pipeline, the random forest regression was chosen as the classical model of choice since it outperformed other models in the workflow regardless of the training data. Our main results on the performance of classical and QRC-enhanced pipelines are summarized in Fig.~\ref{fig:combined-model-metric-figure}~\footnote{800 record samples for MMACD 5, 9, 14, and 15 features were chosen based on SHAP gradient method as compared to MMACD 4 which used SHAP kernel method to save on compute resources, as the change is verified to have minimal impact on the results.}. Taking Fig.~\ref{fig:combined-model-metric-figure}(d) as a representative example, the QRC embeddings appear to result in an improvement upon the classical model performance for small training data sizes (e.g., 100 and 200 records). The average MSE is better for the QRC-enhanced models versus purely classical models across an average of five samples for 100 records. When compared to the classical Gaussian RBF kernel, we find that the QRC embedding encodes the data in a manner that outperforms the classical Gaussian RBF kernel in both the entire data sample and the five subsample averages for 100, 200, and 800 records.  In addition, we also observe that QRC embeddings that include two-body observables outperformed one-body-observable embeddings. For larger subsamples consisting of 800 records, the classical input features yielded similar performance as the QRC embedded features across an average of 5 runs. 

\begin{table*}
\caption{\justifying \label{tab:SVM_UMAP_embedding}\textbf{MMACD activity classification performance on 2D UMAP embeddings}. 2D UMAP embedding data for MMACD dataset 4 is converted to binary classification task by high/low median cut. Performance of support vector machine (SVM) on 100, 200, and 800 record subsamples is reported with both QRC and classical embedding features. A five-subsample average is used to estimate the uncertainty as the standard deviation over the sample.}
\begin{ruledtabular}
\begin{tabular}{cccccccc}
&&\multicolumn{2}{c}{100 records}&\multicolumn{2}{c}{200 records}&\multicolumn{2}{c}{800 records}\\
 &Metric&Mean&Std. Dev.&Mean
&Std. Dev.&Mean&Std. Dev.\\ \hline
 Classical&Accuracy&0.660&$\pm 0.095$&0.737&$\pm 0.041$&0.709&$\pm 0.027$ \\
 &Precision&0.672&$\pm 0.115$&0.784&$\pm 0.098$&0.727&$\pm 0.065$\\
 &Recall&0.653&$\pm 0.073$&0.680&$\pm 0.143$&0.689&$\pm 0.077$\\
 &F1 &0.660&$\pm 0.084$&0.716&$\pm 0.061$&0.702&$\pm 0.019$\\
 QRC&Accuracy&0.747&$\pm 0.077$&0.760&$\pm 0.042$&0.742&$\pm 0.033$ \\
 &Precision&0.766&$\pm 0.068$&0.776&$\pm 0.061$&0.763&$\pm 0.063$\\
 &Recall&0.707&$\pm 0.101$&0.747&$\pm 0.139$&0.713&$\pm 0.062$\\
 &F1 &0.734&$\pm 0.085$&0.752&$\pm 0.062$&0.734&$\pm 0.028$\\
\end{tabular}
\end{ruledtabular}
\end{table*}

Furthermore, Fig.~\ref{fig:combined-model-metric-figure} also reveals a consistent trend where QRC embeddings perform equally or better than classical methods across the five different MMACD data sets. QRC two-body observable embeddings generally match or outperform QRC one-body observable embeddings across these five data sets as well. In contrast, the 800 data records results consistently show that QRC embedded data no longer has the advantage at this sample size. We note, however, that the QRC performance advantage size is generically comparable to the performance error estimates. This is, in large part, a direct consequence of the small sample size of the datasets considered. To check the robustness of our results against limited dataset subsampling, we also ran simulations with 25 subsamples of 100 records for MMACD 14, as shown in Fig.~\ref{fig:act14-25subsamp}. While the performance estimates and the sizes of error bars for classical methods and the one-body embedding QRC remain similar to the 5 subsample case, the two-body QRC embedding performance estimate improves and becomes more certain. As a result, we observe QRC performance improvements over classical methods that are outside of the error estimates in spite of the small dataset size. Overall, the consistent QRC performance improvements that we observe across multiple molecular activity datasets at small sample sizes indicate that the feature engineering process utilizing QRC techniques could help improve the stability of the classical machine learning model, reducing the need for large training datasets. This could make QRC potentially valuable in application areas involving small datasets, which are common in chemistry and health applications where data collection is expensive and complex.

\subsection{UMAP results}\label{sec:umapres}

In this section, we present the findings from our UMAP analysis. UMAP was utilized to project high-dimensional data into a two-dimensional space, enabling the visualization and identification of clusters. This dimensionality reduction technique preserves the local and global structure of the data~\cite{mcinnes2020umapuniformmanifoldapproximation, Dorrity2020}, facilitating the interpretation of data patterns and relationships.  We summarize representative results for the case of 200-sized subsampled datasets derived from MMACD 4 in Fig.~\ref{fig:umap_200recs}. Although there is a clear concentration of high activity values for raw classical features, the correlation of data geometry and molecular activity is manifestly improved for QRC embeddings. The QRC one-body-observable embedding shows a similar pattern to the regular QRC embedding but with a tighter clustering of low ACT values. QRC data clustering in low dimensions also has manifest advantages when compared with the classical non-linear Gaussian kernel. Thus, we hypothesize that the improved QRC clustering in the low dimensional space is not purely an effect of the QRC kernel non-linearity but is an intrinsic feature of the QRC embeddings. 

The distinct clustering patterns observed in the UMAP visualization (Fig.~\ref{fig:umap_200recs}) after SHAP-based feature selection reveal fundamental differences in how classical and quantum methods process molecular information. The scattered distribution in the classical approach indicates a more continuous representation of molecular features, suggesting linear or gradual transitions between different molecular activities. In contrast, the emergence of two clear clusters in the QRC visualization suggests that the quantum processing has identified distinct underlying patterns in the molecular activity space, potentially leveraging quantum effects to better distinguish between different molecular behavior classes. This enhanced clustering capability of QRC on low dimensional UMAP projection suggests the potential to capture complex, nonlinear relationships between molecular features with a more interpretable and therefore also more robust model. 

In order to quantify the relationship between the interpretability in UMAP and the model performance, we used the 2D UMAP manifold embeddings in a readily interpretable binary classification task. The classification task was derived from the original by splitting the target activity variable at the median into high and low-value classes. The classical and QRC UMAP embeddings were input for a Support Vector Machine, whose performance we calculate from the five-run average for the 100, 200, and 800 record subsamples. As can be seen in Table \ref{tab:SVM_UMAP_embedding}, the QRC UMAP embedding outperforms the classical embedding for all record sizes, thus quantifying the difference between UMAP embeddings in Fig.~\ref{fig:umap_200recs}.  However, we note again that the performance improvements are comparable to the error estimates, in large part due to the small sample sizes in the dataset. We conjecture that the improved interpretability of the QRC embeddings in low dimensions might lead to the better performance of the QRC method for small sample sizes observed throughout MMACD datasets previously. While the abundance of data can overcome suboptimal data clustering in a high-dimensional data representation, the interpretability of the data in terms of simpler low-dimensional structures might lead to performance gains when data is scarce.

\section{Conclusions and outlook}\label{sec:conclusions} 
In conclusion, our study presented a systematic exploration of QRC method on biomedical data exemplified by MMACD. We investigated the performance of both classical and QRC models at different sample sizes as well as the interpretability of the data at low feature dimensions using the UMAP technique. Our results suggest that QRC embeddings result in comparatively more robust, stable model performance for smaller datasets. The stability in QRC performance for small training data is further augmented by the manifestly more interpretable nature of QRC-derived features when projected by UMAP to low-dimensional feature space. Consequently, this study suggests a potential for QRC-enhanced models in the domain of biomedical data science for use cases that necessitate small training set sizes and readily interpretable, robust models. 

We believe the results presented in this paper will motivate future studies across several directions. Firstly, the use of quantum hardware could allow for effective QRC embedding generation for more classical input features than the cost of numerical simulations allows for -- in fact, recent QRC experiments have been effective at 100 feature (qubit) scale~\cite{kornjaca2024}. Further increase in qubit counts and improved coherence times could enable QRC to handle larger datasets and more complex molecular systems, facilitating more extensive comparisons with classical methods. In fact, such extensions to more varied and feature-rich datasets could be a route to overcome the main caveat of our results and the large uncertainty estimates that result from modeling on small dataset sizes.  Furthermore, a wide range of alternative feature selection techniques, such as genetic algorithms, could be relevant for selecting relevant molecular descriptors for the QRC pipeline and validating their biological relevance. Alternatively, a QRC pre-processing could help in learning the correlations using Bayesian network learning methods~\cite{Sachs_Perez_Pe’er_Lauffenburger_Nolan_2005, Gao2022}. Finally, collaborative efforts with experimental laboratories will be crucial for empirically validating QRC's computational predictions, bridging the gap between theoretical advantages and practical clinical applications~\cite{vakili2024quantumcomputingenhancedalgorithmunveils}.

\section*{Acknowledgements}
We would like to thank Gopal Karemore, Scott Buchholz, Fangli Liu, Yuval Boger, and Chen Zhao for the fruitful discussion and support during this research project. 

 % Produces the bibliography via BibTeX.
\bibliography{qrc_robustness}

\begin{thebibliography}{69}%
\makeatletter
\providecommand \@ifxundefined [1]{%
 \@ifx{#1\undefined}
}%
\providecommand \@ifnum [1]{%
 \ifnum #1\expandafter \@firstoftwo
 \else \expandafter \@secondoftwo
 \fi
}%
\providecommand \@ifx [1]{%
 \ifx #1\expandafter \@firstoftwo
 \else \expandafter \@secondoftwo
 \fi
}%
\providecommand \natexlab [1]{#1}%
\providecommand \enquote  [1]{``#1''}%
\providecommand \bibnamefont  [1]{#1}%
\providecommand \bibfnamefont [1]{#1}%
\providecommand \citenamefont [1]{#1}%
\providecommand \href@noop [0]{\@secondoftwo}%
\providecommand \href [0]{\begingroup \@sanitize@url \@href}%
\providecommand \@href[1]{\@@startlink{#1}\@@href}%
\providecommand \@@href[1]{\endgroup#1\@@endlink}%
\providecommand \@sanitize@url [0]{\catcode `\\12\catcode `\$12\catcode `\&12\catcode `\#12\catcode `\^12\catcode `\_12\catcode `\%12\relax}%
\providecommand \@@startlink[1]{}%
\providecommand \@@endlink[0]{}%
\providecommand \url  [0]{\begingroup\@sanitize@url \@url }%
\providecommand \@url [1]{\endgroup\@href {#1}{\urlprefix }}%
\providecommand \urlprefix  [0]{URL }%
\providecommand \Eprint [0]{\href }%
\providecommand \doibase [0]{http://dx.doi.org/}%
\providecommand \selectlanguage [0]{\@gobble}%
\providecommand \bibinfo  [0]{\@secondoftwo}%
\providecommand \bibfield  [0]{\@secondoftwo}%
\providecommand \translation [1]{[#1]}%
\providecommand \BibitemOpen [0]{}%
\providecommand \bibitemStop [0]{}%
\providecommand \bibitemNoStop [0]{.\EOS\space}%
\providecommand \EOS [0]{\spacefactor3000\relax}%
\providecommand \BibitemShut  [1]{\csname bibitem#1\endcsname}%
\let\auto@bib@innerbib\@empty
%</preamble>
\bibitem [{\citenamefont {Liu}\ \emph {et~al.}(2018)\citenamefont {Liu}, \citenamefont {Wang}, \citenamefont {Glover}, \citenamefont {Feasel},\ and\ \citenamefont {Wallqvist}}]{liu2016}%
  \BibitemOpen
  \bibfield  {author} {\bibinfo {author} {\bibfnamefont {R.}~\bibnamefont {Liu}}, \bibinfo {author} {\bibfnamefont {H.}~\bibnamefont {Wang}}, \bibinfo {author} {\bibfnamefont {K.}~\bibnamefont {Glover}}, \bibinfo {author} {\bibfnamefont {M.}~\bibnamefont {Feasel}}, \ and\ \bibinfo {author} {\bibfnamefont {A.}~\bibnamefont {Wallqvist}},\ }\href {\doibase 10.1021/acs.jcim.8b00348} {\bibfield  {journal} {\bibinfo  {journal} {Journal of Chemical Information and Modeling}\ }\textbf {\bibinfo {volume} {59}} (\bibinfo {year} {2018}),\ 10.1021/acs.jcim.8b00348}\BibitemShut {NoStop}%
\bibitem [{\citenamefont {Pantaleão}\ \emph {et~al.}(2022)\citenamefont {Pantaleão}, \citenamefont {Fernandes}, \citenamefont {Gonçalves}, \citenamefont {Maltarollo},\ and\ \citenamefont {Honorio}}]{pharmacokinetics_review}%
  \BibitemOpen
  \bibfield  {author} {\bibinfo {author} {\bibfnamefont {S.~Q.}\ \bibnamefont {Pantaleão}}, \bibinfo {author} {\bibfnamefont {P.~O.}\ \bibnamefont {Fernandes}}, \bibinfo {author} {\bibfnamefont {J.~E.}\ \bibnamefont {Gonçalves}}, \bibinfo {author} {\bibfnamefont {V.~G.}\ \bibnamefont {Maltarollo}}, \ and\ \bibinfo {author} {\bibfnamefont {K.~M.}\ \bibnamefont {Honorio}},\ }\href {\doibase https://doi.org/10.1002/cmdc.202100542} {\bibfield  {journal} {\bibinfo  {journal} {ChemMedChem}\ }\textbf {\bibinfo {volume} {17}},\ \bibinfo {pages} {e202100542} (\bibinfo {year} {2022})}\BibitemShut {NoStop}%
\bibitem [{\citenamefont {Vamathevan}\ \emph {et~al.}(2019)\citenamefont {Vamathevan}, \citenamefont {Clark}, \citenamefont {Czodrowski}, \citenamefont {Dunham}, \citenamefont {Ferran}, \citenamefont {Lee}, \citenamefont {Li}, \citenamefont {Madabhushi}, \citenamefont {Shah}, \citenamefont {Spitzer},\ and\ \citenamefont {Zhao}}]{vamathevan2019}%
  \BibitemOpen
  \bibfield  {author} {\bibinfo {author} {\bibfnamefont {J.}~\bibnamefont {Vamathevan}}, \bibinfo {author} {\bibfnamefont {D.}~\bibnamefont {Clark}}, \bibinfo {author} {\bibfnamefont {P.}~\bibnamefont {Czodrowski}}, \bibinfo {author} {\bibfnamefont {I.}~\bibnamefont {Dunham}}, \bibinfo {author} {\bibfnamefont {E.}~\bibnamefont {Ferran}}, \bibinfo {author} {\bibfnamefont {G.}~\bibnamefont {Lee}}, \bibinfo {author} {\bibfnamefont {B.}~\bibnamefont {Li}}, \bibinfo {author} {\bibfnamefont {A.}~\bibnamefont {Madabhushi}}, \bibinfo {author} {\bibfnamefont {P.}~\bibnamefont {Shah}}, \bibinfo {author} {\bibfnamefont {M.}~\bibnamefont {Spitzer}}, \ and\ \bibinfo {author} {\bibfnamefont {S.}~\bibnamefont {Zhao}},\ }\href {\doibase 10.1038/s41573-019-0024-5} {\bibfield  {journal} {\bibinfo  {journal} {Nature Reviews Drug Discovery}\ }\textbf {\bibinfo {volume} {18}},\ \bibinfo {pages} {463} (\bibinfo {year} {2019})}\BibitemShut {NoStop}%
\bibitem [{\citenamefont {Deng}\ \emph {et~al.}(2023)\citenamefont {Deng}, \citenamefont {Yang}, \citenamefont {Wang}, \citenamefont {Ojima}, \citenamefont {Samaras},\ and\ \citenamefont {Wang}}]{Deng2023}%
  \BibitemOpen
  \bibfield  {author} {\bibinfo {author} {\bibfnamefont {J.}~\bibnamefont {Deng}}, \bibinfo {author} {\bibfnamefont {Z.}~\bibnamefont {Yang}}, \bibinfo {author} {\bibfnamefont {H.}~\bibnamefont {Wang}}, \bibinfo {author} {\bibfnamefont {I.}~\bibnamefont {Ojima}}, \bibinfo {author} {\bibfnamefont {D.}~\bibnamefont {Samaras}}, \ and\ \bibinfo {author} {\bibfnamefont {F.}~\bibnamefont {Wang}},\ }\href {\doibase 10.1038/s41467-023-41948-6} {\bibfield  {journal} {\bibinfo  {journal} {Nature Communications}\ }\textbf {\bibinfo {volume} {14}},\ \bibinfo {pages} {6395} (\bibinfo {year} {2023})}\BibitemShut {NoStop}%
\bibitem [{\citenamefont {Zeng}\ \emph {et~al.}(2022)\citenamefont {Zeng}, \citenamefont {Xiang}, \citenamefont {Yu}, \citenamefont {Wang}, \citenamefont {Li}, \citenamefont {Nussinov},\ and\ \citenamefont {Cheng}}]{Zeng2022}%
  \BibitemOpen
  \bibfield  {author} {\bibinfo {author} {\bibfnamefont {X.}~\bibnamefont {Zeng}}, \bibinfo {author} {\bibfnamefont {H.}~\bibnamefont {Xiang}}, \bibinfo {author} {\bibfnamefont {L.}~\bibnamefont {Yu}}, \bibinfo {author} {\bibfnamefont {J.}~\bibnamefont {Wang}}, \bibinfo {author} {\bibfnamefont {K.}~\bibnamefont {Li}}, \bibinfo {author} {\bibfnamefont {R.}~\bibnamefont {Nussinov}}, \ and\ \bibinfo {author} {\bibfnamefont {F.}~\bibnamefont {Cheng}},\ }\href {\doibase 10.1038/s42256-022-00557-6} {\bibfield  {journal} {\bibinfo  {journal} {Nature Machine Intelligence}\ }\textbf {\bibinfo {volume} {4}},\ \bibinfo {pages} {1004} (\bibinfo {year} {2022})}\BibitemShut {NoStop}%
\bibitem [{\citenamefont {Team}({\natexlab{a}})}]{rdkit}%
  \BibitemOpen
  \bibfield  {author} {\bibinfo {author} {\bibfnamefont {R.}~\bibnamefont {Team}},\ }\href@noop {} {\enquote {\bibinfo {title} {Rdkit: Open-source cheminformatics},}\ }\bibinfo {howpublished} {\url{http://www.rdkit.org}} ({\natexlab{a}}),\ \bibinfo {note} {accessed: 2024-06-04}\BibitemShut {NoStop}%
\bibitem [{\citenamefont {Comesana}\ \emph {et~al.}(2022)\citenamefont {Comesana}, \citenamefont {Huntington}, \citenamefont {Scown}, \citenamefont {Niemeyer},\ and\ \citenamefont {Rapp}}]{COMESANA2022123836}%
  \BibitemOpen
  \bibfield  {author} {\bibinfo {author} {\bibfnamefont {A.~E.}\ \bibnamefont {Comesana}}, \bibinfo {author} {\bibfnamefont {T.~T.}\ \bibnamefont {Huntington}}, \bibinfo {author} {\bibfnamefont {C.~D.}\ \bibnamefont {Scown}}, \bibinfo {author} {\bibfnamefont {K.~E.}\ \bibnamefont {Niemeyer}}, \ and\ \bibinfo {author} {\bibfnamefont {V.~H.}\ \bibnamefont {Rapp}},\ }\href {\doibase https://doi.org/10.1016/j.fuel.2022.123836} {\bibfield  {journal} {\bibinfo  {journal} {Fuel}\ }\textbf {\bibinfo {volume} {321}},\ \bibinfo {pages} {123836} (\bibinfo {year} {2022})}\BibitemShut {NoStop}%
\bibitem [{\citenamefont {Cereto-Massagué}\ \emph {et~al.}(2015)\citenamefont {Cereto-Massagué}, \citenamefont {Ojeda}, \citenamefont {Valls}, \citenamefont {Mulero}, \citenamefont {Garcia-Vallvé},\ and\ \citenamefont {Pujadas}}]{CERETOMASSAGUE201558}%
  \BibitemOpen
  \bibfield  {author} {\bibinfo {author} {\bibfnamefont {A.}~\bibnamefont {Cereto-Massagué}}, \bibinfo {author} {\bibfnamefont {M.~J.}\ \bibnamefont {Ojeda}}, \bibinfo {author} {\bibfnamefont {C.}~\bibnamefont {Valls}}, \bibinfo {author} {\bibfnamefont {M.}~\bibnamefont {Mulero}}, \bibinfo {author} {\bibfnamefont {S.}~\bibnamefont {Garcia-Vallvé}}, \ and\ \bibinfo {author} {\bibfnamefont {G.}~\bibnamefont {Pujadas}},\ }\href {\doibase https://doi.org/10.1016/j.ymeth.2014.08.005} {\bibfield  {journal} {\bibinfo  {journal} {Methods}\ }\textbf {\bibinfo {volume} {71}},\ \bibinfo {pages} {58} (\bibinfo {year} {2015})},\ \bibinfo {note} {virtual Screening}\BibitemShut {NoStop}%
\bibitem [{\citenamefont {Mellor}\ \emph {et~al.}(2019)\citenamefont {Mellor}, \citenamefont {{Marchese Robinson}}, \citenamefont {Benigni}, \citenamefont {Ebbrell}, \citenamefont {Enoch}, \citenamefont {Firman}, \citenamefont {Madden}, \citenamefont {Pawar}, \citenamefont {Yang},\ and\ \citenamefont {Cronin}}]{MELLOR2019121}%
  \BibitemOpen
  \bibfield  {author} {\bibinfo {author} {\bibfnamefont {C.}~\bibnamefont {Mellor}}, \bibinfo {author} {\bibfnamefont {R.}~\bibnamefont {{Marchese Robinson}}}, \bibinfo {author} {\bibfnamefont {R.}~\bibnamefont {Benigni}}, \bibinfo {author} {\bibfnamefont {D.}~\bibnamefont {Ebbrell}}, \bibinfo {author} {\bibfnamefont {S.}~\bibnamefont {Enoch}}, \bibinfo {author} {\bibfnamefont {J.}~\bibnamefont {Firman}}, \bibinfo {author} {\bibfnamefont {J.}~\bibnamefont {Madden}}, \bibinfo {author} {\bibfnamefont {G.}~\bibnamefont {Pawar}}, \bibinfo {author} {\bibfnamefont {C.}~\bibnamefont {Yang}}, \ and\ \bibinfo {author} {\bibfnamefont {M.}~\bibnamefont {Cronin}},\ }\href {\doibase https://doi.org/10.1016/j.yrtph.2018.11.002} {\bibfield  {journal} {\bibinfo  {journal} {Regulatory Toxicology and Pharmacology}\ }\textbf {\bibinfo {volume} {101}},\ \bibinfo {pages} {121} (\bibinfo {year} {2019})}\BibitemShut {NoStop}%
\bibitem [{\citenamefont {Vu}\ \emph {et~al.}(2023)\citenamefont {Vu}, \citenamefont {Ha}, \citenamefont {Nguyen}, \citenamefont {Nguyen}, \citenamefont {Abe}, \citenamefont {Tran}, \citenamefont {Tran}, \citenamefont {Kino}, \citenamefont {Miyake}, \citenamefont {Tsuda},\ and\ \citenamefont {Dam}}]{Vu2023}%
  \BibitemOpen
  \bibfield  {author} {\bibinfo {author} {\bibfnamefont {T.-S.}\ \bibnamefont {Vu}}, \bibinfo {author} {\bibfnamefont {M.-Q.}\ \bibnamefont {Ha}}, \bibinfo {author} {\bibfnamefont {D.-N.}\ \bibnamefont {Nguyen}}, \bibinfo {author} {\bibfnamefont {V.-C.}\ \bibnamefont {Nguyen}}, \bibinfo {author} {\bibfnamefont {Y.}~\bibnamefont {Abe}}, \bibinfo {author} {\bibfnamefont {T.}~\bibnamefont {Tran}}, \bibinfo {author} {\bibfnamefont {H.}~\bibnamefont {Tran}}, \bibinfo {author} {\bibfnamefont {H.}~\bibnamefont {Kino}}, \bibinfo {author} {\bibfnamefont {T.}~\bibnamefont {Miyake}}, \bibinfo {author} {\bibfnamefont {K.}~\bibnamefont {Tsuda}}, \ and\ \bibinfo {author} {\bibfnamefont {H.-C.}\ \bibnamefont {Dam}},\ }\href {\doibase 10.1038/s41524-023-01163-9} {\bibfield  {journal} {\bibinfo  {journal} {npj Computational Materials}\ }\textbf {\bibinfo {volume} {9}},\ \bibinfo {pages} {215} (\bibinfo {year} {2023})}\BibitemShut {NoStop}%
\bibitem [{\citenamefont {Caro}\ \emph {et~al.}(2022)\citenamefont {Caro}, \citenamefont {Huang}, \citenamefont {Cerezo}, \citenamefont {Sharma}, \citenamefont {Sornborger}, \citenamefont {Cincio},\ and\ \citenamefont {Coles}}]{Caro_2022}%
  \BibitemOpen
  \bibfield  {author} {\bibinfo {author} {\bibfnamefont {M.~C.}\ \bibnamefont {Caro}}, \bibinfo {author} {\bibfnamefont {H.-Y.}\ \bibnamefont {Huang}}, \bibinfo {author} {\bibfnamefont {M.}~\bibnamefont {Cerezo}}, \bibinfo {author} {\bibfnamefont {K.}~\bibnamefont {Sharma}}, \bibinfo {author} {\bibfnamefont {A.}~\bibnamefont {Sornborger}}, \bibinfo {author} {\bibfnamefont {L.}~\bibnamefont {Cincio}}, \ and\ \bibinfo {author} {\bibfnamefont {P.~J.}\ \bibnamefont {Coles}},\ }\href {\doibase 10.1038/s41467-022-32550-3} {\bibfield  {journal} {\bibinfo  {journal} {Nature Communications}\ }\textbf {\bibinfo {volume} {13}} (\bibinfo {year} {2022}),\ 10.1038/s41467-022-32550-3}\BibitemShut {NoStop}%
\bibitem [{\citenamefont {Schuld}\ \emph {et~al.}(2014)\citenamefont {Schuld}, \citenamefont {Sinayskiy},\ and\ \citenamefont {Petruccione}}]{Schuld_2014}%
  \BibitemOpen
  \bibfield  {author} {\bibinfo {author} {\bibfnamefont {M.}~\bibnamefont {Schuld}}, \bibinfo {author} {\bibfnamefont {I.}~\bibnamefont {Sinayskiy}}, \ and\ \bibinfo {author} {\bibfnamefont {F.}~\bibnamefont {Petruccione}},\ }\href {\doibase 10.1080/00107514.2014.964942} {\bibfield  {journal} {\bibinfo  {journal} {Contemporary Physics}\ }\textbf {\bibinfo {volume} {56}},\ \bibinfo {pages} {172–185} (\bibinfo {year} {2014})}\BibitemShut {NoStop}%
\bibitem [{\citenamefont {Carleo}\ \emph {et~al.}(2019)\citenamefont {Carleo}, \citenamefont {Cirac}, \citenamefont {Cranmer}, \citenamefont {Daudet}, \citenamefont {Schuld}, \citenamefont {Tishby}, \citenamefont {Vogt-Maranto},\ and\ \citenamefont {Zdeborová}}]{Carleo_2019}%
  \BibitemOpen
  \bibfield  {author} {\bibinfo {author} {\bibfnamefont {G.}~\bibnamefont {Carleo}}, \bibinfo {author} {\bibfnamefont {I.}~\bibnamefont {Cirac}}, \bibinfo {author} {\bibfnamefont {K.}~\bibnamefont {Cranmer}}, \bibinfo {author} {\bibfnamefont {L.}~\bibnamefont {Daudet}}, \bibinfo {author} {\bibfnamefont {M.}~\bibnamefont {Schuld}}, \bibinfo {author} {\bibfnamefont {N.}~\bibnamefont {Tishby}}, \bibinfo {author} {\bibfnamefont {L.}~\bibnamefont {Vogt-Maranto}}, \ and\ \bibinfo {author} {\bibfnamefont {L.}~\bibnamefont {Zdeborová}},\ }\href {\doibase 10.1103/revmodphys.91.045002} {\bibfield  {journal} {\bibinfo  {journal} {Reviews of Modern Physics}\ }\textbf {\bibinfo {volume} {91}} (\bibinfo {year} {2019}),\ 10.1103/revmodphys.91.045002}\BibitemShut {NoStop}%
\bibitem [{\citenamefont {Biamonte}\ \emph {et~al.}(2017)\citenamefont {Biamonte}, \citenamefont {Wittek}, \citenamefont {Pancotti}, \citenamefont {Rebentrost}, \citenamefont {Wiebe},\ and\ \citenamefont {Lloyd}}]{Biamonte_2017}%
  \BibitemOpen
  \bibfield  {author} {\bibinfo {author} {\bibfnamefont {J.}~\bibnamefont {Biamonte}}, \bibinfo {author} {\bibfnamefont {P.}~\bibnamefont {Wittek}}, \bibinfo {author} {\bibfnamefont {N.}~\bibnamefont {Pancotti}}, \bibinfo {author} {\bibfnamefont {P.}~\bibnamefont {Rebentrost}}, \bibinfo {author} {\bibfnamefont {N.}~\bibnamefont {Wiebe}}, \ and\ \bibinfo {author} {\bibfnamefont {S.}~\bibnamefont {Lloyd}},\ }\href {\doibase 10.1038/nature23474} {\bibfield  {journal} {\bibinfo  {journal} {Nature}\ }\textbf {\bibinfo {volume} {549}},\ \bibinfo {pages} {195–202} (\bibinfo {year} {2017})}\BibitemShut {NoStop}%
\bibitem [{\citenamefont {Lloyd}\ \emph {et~al.}(2020)\citenamefont {Lloyd}, \citenamefont {Schuld}, \citenamefont {Ijaz}, \citenamefont {Izaac},\ and\ \citenamefont {Killoran}}]{lloyd2020quantumembeddingsmachinelearning}%
  \BibitemOpen
  \bibfield  {author} {\bibinfo {author} {\bibfnamefont {S.}~\bibnamefont {Lloyd}}, \bibinfo {author} {\bibfnamefont {M.}~\bibnamefont {Schuld}}, \bibinfo {author} {\bibfnamefont {A.}~\bibnamefont {Ijaz}}, \bibinfo {author} {\bibfnamefont {J.}~\bibnamefont {Izaac}}, \ and\ \bibinfo {author} {\bibfnamefont {N.}~\bibnamefont {Killoran}},\ }\href {https://arxiv.org/abs/2001.03622} {\enquote {\bibinfo {title} {Quantum embeddings for machine learning},}\ } (\bibinfo {year} {2020}),\ \Eprint {http://arxiv.org/abs/2001.03622} {arXiv:2001.03622 [quant-ph]} \BibitemShut {NoStop}%
\bibitem [{\citenamefont {Schuld}(2021)}]{schuld2021supervisedquantummachinelearning}%
  \BibitemOpen
  \bibfield  {author} {\bibinfo {author} {\bibfnamefont {M.}~\bibnamefont {Schuld}},\ }\href {https://arxiv.org/abs/2101.11020} {\enquote {\bibinfo {title} {Supervised quantum machine learning models are kernel methods},}\ } (\bibinfo {year} {2021}),\ \Eprint {http://arxiv.org/abs/2101.11020} {arXiv:2101.11020 [quant-ph]} \BibitemShut {NoStop}%
\bibitem [{\citenamefont {Killoran}\ \emph {et~al.}(2019)\citenamefont {Killoran}, \citenamefont {Bromley}, \citenamefont {Arrazola}, \citenamefont {Schuld}, \citenamefont {Quesada},\ and\ \citenamefont {Lloyd}}]{Killoran_2019}%
  \BibitemOpen
  \bibfield  {author} {\bibinfo {author} {\bibfnamefont {N.}~\bibnamefont {Killoran}}, \bibinfo {author} {\bibfnamefont {T.~R.}\ \bibnamefont {Bromley}}, \bibinfo {author} {\bibfnamefont {J.~M.}\ \bibnamefont {Arrazola}}, \bibinfo {author} {\bibfnamefont {M.}~\bibnamefont {Schuld}}, \bibinfo {author} {\bibfnamefont {N.}~\bibnamefont {Quesada}}, \ and\ \bibinfo {author} {\bibfnamefont {S.}~\bibnamefont {Lloyd}},\ }\href {\doibase 10.1103/physrevresearch.1.033063} {\bibfield  {journal} {\bibinfo  {journal} {Physical Review Research}\ }\textbf {\bibinfo {volume} {1}} (\bibinfo {year} {2019}),\ 10.1103/physrevresearch.1.033063}\BibitemShut {NoStop}%
\bibitem [{\citenamefont {Huembeli}\ \emph {et~al.}(2022)\citenamefont {Huembeli}, \citenamefont {Carleo},\ and\ \citenamefont {Mezzacapo}}]{huembeli2022entanglementforginggenerativeneural}%
  \BibitemOpen
  \bibfield  {author} {\bibinfo {author} {\bibfnamefont {P.}~\bibnamefont {Huembeli}}, \bibinfo {author} {\bibfnamefont {G.}~\bibnamefont {Carleo}}, \ and\ \bibinfo {author} {\bibfnamefont {A.}~\bibnamefont {Mezzacapo}},\ }\href {https://arxiv.org/abs/2205.00933} {\enquote {\bibinfo {title} {Entanglement forging with generative neural network models},}\ } (\bibinfo {year} {2022}),\ \Eprint {http://arxiv.org/abs/2205.00933} {arXiv:2205.00933 [quant-ph]} \BibitemShut {NoStop}%
\bibitem [{\citenamefont {Havlíček}\ \emph {et~al.}(2019)\citenamefont {Havlíček}, \citenamefont {Córcoles}, \citenamefont {Temme}, \citenamefont {Harrow}, \citenamefont {Kandala}, \citenamefont {Chow},\ and\ \citenamefont {Gambetta}}]{Havl_ek_2019}%
  \BibitemOpen
  \bibfield  {author} {\bibinfo {author} {\bibfnamefont {V.}~\bibnamefont {Havlíček}}, \bibinfo {author} {\bibfnamefont {A.~D.}\ \bibnamefont {Córcoles}}, \bibinfo {author} {\bibfnamefont {K.}~\bibnamefont {Temme}}, \bibinfo {author} {\bibfnamefont {A.~W.}\ \bibnamefont {Harrow}}, \bibinfo {author} {\bibfnamefont {A.}~\bibnamefont {Kandala}}, \bibinfo {author} {\bibfnamefont {J.~M.}\ \bibnamefont {Chow}}, \ and\ \bibinfo {author} {\bibfnamefont {J.~M.}\ \bibnamefont {Gambetta}},\ }\href {\doibase 10.1038/s41586-019-0980-2} {\bibfield  {journal} {\bibinfo  {journal} {Nature}\ }\textbf {\bibinfo {volume} {567}},\ \bibinfo {pages} {209–212} (\bibinfo {year} {2019})}\BibitemShut {NoStop}%
\bibitem [{\citenamefont {Schuld}\ and\ \citenamefont {Killoran}(2019)}]{Schuld_2019}%
  \BibitemOpen
  \bibfield  {author} {\bibinfo {author} {\bibfnamefont {M.}~\bibnamefont {Schuld}}\ and\ \bibinfo {author} {\bibfnamefont {N.}~\bibnamefont {Killoran}},\ }\href {\doibase 10.1103/physrevlett.122.040504} {\bibfield  {journal} {\bibinfo  {journal} {Physical Review Letters}\ }\textbf {\bibinfo {volume} {122}} (\bibinfo {year} {2019}),\ 10.1103/physrevlett.122.040504}\BibitemShut {NoStop}%
\bibitem [{\citenamefont {Liu}\ \emph {et~al.}(2021)\citenamefont {Liu}, \citenamefont {Arunachalam},\ and\ \citenamefont {Temme}}]{Liu_2021}%
  \BibitemOpen
  \bibfield  {author} {\bibinfo {author} {\bibfnamefont {Y.}~\bibnamefont {Liu}}, \bibinfo {author} {\bibfnamefont {S.}~\bibnamefont {Arunachalam}}, \ and\ \bibinfo {author} {\bibfnamefont {K.}~\bibnamefont {Temme}},\ }\href {\doibase 10.1038/s41567-021-01287-z} {\bibfield  {journal} {\bibinfo  {journal} {Nature Physics}\ }\textbf {\bibinfo {volume} {17}},\ \bibinfo {pages} {1013–1017} (\bibinfo {year} {2021})}\BibitemShut {NoStop}%
\bibitem [{\citenamefont {Huang}\ \emph {et~al.}(2021)\citenamefont {Huang}, \citenamefont {Broughton}, \citenamefont {Mohseni}, \citenamefont {Babbush}, \citenamefont {Boixo}, \citenamefont {Neven},\ and\ \citenamefont {McClean}}]{Huang_2021}%
  \BibitemOpen
  \bibfield  {author} {\bibinfo {author} {\bibfnamefont {H.-Y.}\ \bibnamefont {Huang}}, \bibinfo {author} {\bibfnamefont {M.}~\bibnamefont {Broughton}}, \bibinfo {author} {\bibfnamefont {M.}~\bibnamefont {Mohseni}}, \bibinfo {author} {\bibfnamefont {R.}~\bibnamefont {Babbush}}, \bibinfo {author} {\bibfnamefont {S.}~\bibnamefont {Boixo}}, \bibinfo {author} {\bibfnamefont {H.}~\bibnamefont {Neven}}, \ and\ \bibinfo {author} {\bibfnamefont {J.~R.}\ \bibnamefont {McClean}},\ }\href {\doibase 10.1038/s41467-021-22539-9} {\bibfield  {journal} {\bibinfo  {journal} {Nature Communications}\ }\textbf {\bibinfo {volume} {12}} (\bibinfo {year} {2021}),\ 10.1038/s41467-021-22539-9}\BibitemShut {NoStop}%
\bibitem [{\citenamefont {Kornjača}\ \emph {et~al.}(2024)\citenamefont {Kornjača}, \citenamefont {Hu}, \citenamefont {Zhao}, \citenamefont {Wurtz}, \citenamefont {Weinberg}, \citenamefont {Hamdan}, \citenamefont {Zhdanov}, \citenamefont {Cantu}, \citenamefont {Zhou}, \citenamefont {Bravo}, \citenamefont {Bagnall}, \citenamefont {Basham}, \citenamefont {Campo}, \citenamefont {Choukri}, \citenamefont {DeAngelo}, \citenamefont {Frederick}, \citenamefont {Haines}, \citenamefont {Hammett}, \citenamefont {Hsu}, \citenamefont {Hu}, \citenamefont {Huber}, \citenamefont {Jepsen}, \citenamefont {Jia}, \citenamefont {Karolyshyn}, \citenamefont {Kwon}, \citenamefont {Long}, \citenamefont {Lopatin}, \citenamefont {Lukin}, \citenamefont {Macrì}, \citenamefont {Marković}, \citenamefont {Martínez-Martínez}, \citenamefont {Meng}, \citenamefont {Ostroumov}, \citenamefont {Paquette}, \citenamefont {Robinson}, \citenamefont {Rodriguez}, \citenamefont {Singh}, \citenamefont {Sinha}, \citenamefont {Thoreen}, \citenamefont
  {Wan}, \citenamefont {Waxman-Lenz}, \citenamefont {Wong}, \citenamefont {Wu}, \citenamefont {Lopes}, \citenamefont {Boger}, \citenamefont {Gemelke}, \citenamefont {Kitagawa}, \citenamefont {Keesling}, \citenamefont {Gao}, \citenamefont {Bylinskii}, \citenamefont {Yelin}, \citenamefont {Liu},\ and\ \citenamefont {Wang}}]{kornjaca2024}%
  \BibitemOpen
  \bibfield  {author} {\bibinfo {author} {\bibfnamefont {M.}~\bibnamefont {Kornjača}}, \bibinfo {author} {\bibfnamefont {H.-Y.}\ \bibnamefont {Hu}}, \bibinfo {author} {\bibfnamefont {C.}~\bibnamefont {Zhao}}, \bibinfo {author} {\bibfnamefont {J.}~\bibnamefont {Wurtz}}, \bibinfo {author} {\bibfnamefont {P.}~\bibnamefont {Weinberg}}, \bibinfo {author} {\bibfnamefont {M.}~\bibnamefont {Hamdan}}, \bibinfo {author} {\bibfnamefont {A.}~\bibnamefont {Zhdanov}}, \bibinfo {author} {\bibfnamefont {S.~H.}\ \bibnamefont {Cantu}}, \bibinfo {author} {\bibfnamefont {H.}~\bibnamefont {Zhou}}, \bibinfo {author} {\bibfnamefont {R.~A.}\ \bibnamefont {Bravo}}, \bibinfo {author} {\bibfnamefont {K.}~\bibnamefont {Bagnall}}, \bibinfo {author} {\bibfnamefont {J.~I.}\ \bibnamefont {Basham}}, \bibinfo {author} {\bibfnamefont {J.}~\bibnamefont {Campo}}, \bibinfo {author} {\bibfnamefont {A.}~\bibnamefont {Choukri}}, \bibinfo {author} {\bibfnamefont {R.}~\bibnamefont {DeAngelo}}, \bibinfo {author} {\bibfnamefont {P.}~\bibnamefont
  {Frederick}}, \bibinfo {author} {\bibfnamefont {D.}~\bibnamefont {Haines}}, \bibinfo {author} {\bibfnamefont {J.}~\bibnamefont {Hammett}}, \bibinfo {author} {\bibfnamefont {N.}~\bibnamefont {Hsu}}, \bibinfo {author} {\bibfnamefont {M.-G.}\ \bibnamefont {Hu}}, \bibinfo {author} {\bibfnamefont {F.}~\bibnamefont {Huber}}, \bibinfo {author} {\bibfnamefont {P.~N.}\ \bibnamefont {Jepsen}}, \bibinfo {author} {\bibfnamefont {N.}~\bibnamefont {Jia}}, \bibinfo {author} {\bibfnamefont {T.}~\bibnamefont {Karolyshyn}}, \bibinfo {author} {\bibfnamefont {M.}~\bibnamefont {Kwon}}, \bibinfo {author} {\bibfnamefont {J.}~\bibnamefont {Long}}, \bibinfo {author} {\bibfnamefont {J.}~\bibnamefont {Lopatin}}, \bibinfo {author} {\bibfnamefont {A.}~\bibnamefont {Lukin}}, \bibinfo {author} {\bibfnamefont {T.}~\bibnamefont {Macrì}}, \bibinfo {author} {\bibfnamefont {O.}~\bibnamefont {Marković}}, \bibinfo {author} {\bibfnamefont {L.~A.}\ \bibnamefont {Martínez-Martínez}}, \bibinfo {author} {\bibfnamefont {X.}~\bibnamefont {Meng}},
  \bibinfo {author} {\bibfnamefont {E.}~\bibnamefont {Ostroumov}}, \bibinfo {author} {\bibfnamefont {D.}~\bibnamefont {Paquette}}, \bibinfo {author} {\bibfnamefont {J.}~\bibnamefont {Robinson}}, \bibinfo {author} {\bibfnamefont {P.~S.}\ \bibnamefont {Rodriguez}}, \bibinfo {author} {\bibfnamefont {A.}~\bibnamefont {Singh}}, \bibinfo {author} {\bibfnamefont {N.}~\bibnamefont {Sinha}}, \bibinfo {author} {\bibfnamefont {H.}~\bibnamefont {Thoreen}}, \bibinfo {author} {\bibfnamefont {N.}~\bibnamefont {Wan}}, \bibinfo {author} {\bibfnamefont {D.}~\bibnamefont {Waxman-Lenz}}, \bibinfo {author} {\bibfnamefont {T.}~\bibnamefont {Wong}}, \bibinfo {author} {\bibfnamefont {K.-H.}\ \bibnamefont {Wu}}, \bibinfo {author} {\bibfnamefont {P.~L.~S.}\ \bibnamefont {Lopes}}, \bibinfo {author} {\bibfnamefont {Y.}~\bibnamefont {Boger}}, \bibinfo {author} {\bibfnamefont {N.}~\bibnamefont {Gemelke}}, \bibinfo {author} {\bibfnamefont {T.}~\bibnamefont {Kitagawa}}, \bibinfo {author} {\bibfnamefont {A.}~\bibnamefont {Keesling}},
  \bibinfo {author} {\bibfnamefont {X.}~\bibnamefont {Gao}}, \bibinfo {author} {\bibfnamefont {A.}~\bibnamefont {Bylinskii}}, \bibinfo {author} {\bibfnamefont {S.~F.}\ \bibnamefont {Yelin}}, \bibinfo {author} {\bibfnamefont {F.}~\bibnamefont {Liu}}, \ and\ \bibinfo {author} {\bibfnamefont {S.-T.}\ \bibnamefont {Wang}},\ }\href {https://arxiv.org/abs/2407.02553} {\enquote {\bibinfo {title} {Large-scale quantum reservoir learning with an analog quantum computer},}\ } (\bibinfo {year} {2024}),\ \Eprint {http://arxiv.org/abs/2407.02553} {arXiv:2407.02553 [quant-ph]} \BibitemShut {NoStop}%
\bibitem [{\citenamefont {Batra}\ \emph {et~al.}(2021)\citenamefont {Batra}, \citenamefont {Zorn}, \citenamefont {Foil}, \citenamefont {Minerali}, \citenamefont {Gawriljuk}, \citenamefont {Lane},\ and\ \citenamefont {Ekins}}]{Batra2021}%
  \BibitemOpen
  \bibfield  {author} {\bibinfo {author} {\bibfnamefont {K.}~\bibnamefont {Batra}}, \bibinfo {author} {\bibfnamefont {K.~M.}\ \bibnamefont {Zorn}}, \bibinfo {author} {\bibfnamefont {D.~H.}\ \bibnamefont {Foil}}, \bibinfo {author} {\bibfnamefont {E.}~\bibnamefont {Minerali}}, \bibinfo {author} {\bibfnamefont {V.~O.}\ \bibnamefont {Gawriljuk}}, \bibinfo {author} {\bibfnamefont {T.~R.}\ \bibnamefont {Lane}}, \ and\ \bibinfo {author} {\bibfnamefont {S.}~\bibnamefont {Ekins}},\ }\href {\doibase 10.1021/acs.jcim.1c00166} {\bibfield  {journal} {\bibinfo  {journal} {Journal of Chemical Information and Modeling}\ }\textbf {\bibinfo {volume} {61}},\ \bibinfo {pages} {2641} (\bibinfo {year} {2021})}\BibitemShut {NoStop}%
\bibitem [{\citenamefont {Bhatia}\ \emph {et~al.}(2023)\citenamefont {Bhatia}, \citenamefont {Saggi},\ and\ \citenamefont {Kais}}]{Bhatia2023}%
  \BibitemOpen
  \bibfield  {author} {\bibinfo {author} {\bibfnamefont {A.~S.}\ \bibnamefont {Bhatia}}, \bibinfo {author} {\bibfnamefont {M.~K.}\ \bibnamefont {Saggi}}, \ and\ \bibinfo {author} {\bibfnamefont {S.}~\bibnamefont {Kais}},\ }\href {\doibase 10.1021/acs.jcim.3c01079} {\bibfield  {journal} {\bibinfo  {journal} {Journal of Chemical Information and Modeling}\ }\textbf {\bibinfo {volume} {63}},\ \bibinfo {pages} {6476} (\bibinfo {year} {2023})}\BibitemShut {NoStop}%
\bibitem [{\citenamefont {Vakili}\ \emph {et~al.}(2024)\citenamefont {Vakili}, \citenamefont {Gorgulla}, \citenamefont {Nigam}, \citenamefont {Bezrukov}, \citenamefont {Varoli}, \citenamefont {Aliper}, \citenamefont {Polykovsky}, \citenamefont {Das}, \citenamefont {Snider}, \citenamefont {Lyakisheva}, \citenamefont {Mansob}, \citenamefont {Yao}, \citenamefont {Bitar}, \citenamefont {Radchenko}, \citenamefont {Ding}, \citenamefont {Liu}, \citenamefont {Meng}, \citenamefont {Ren}, \citenamefont {Cao}, \citenamefont {Stagljar}, \citenamefont {Aspuru-Guzik},\ and\ \citenamefont {Zhavoronkov}}]{vakili2024quantumcomputingenhancedalgorithmunveils}%
  \BibitemOpen
  \bibfield  {author} {\bibinfo {author} {\bibfnamefont {M.~G.}\ \bibnamefont {Vakili}}, \bibinfo {author} {\bibfnamefont {C.}~\bibnamefont {Gorgulla}}, \bibinfo {author} {\bibfnamefont {A.}~\bibnamefont {Nigam}}, \bibinfo {author} {\bibfnamefont {D.}~\bibnamefont {Bezrukov}}, \bibinfo {author} {\bibfnamefont {D.}~\bibnamefont {Varoli}}, \bibinfo {author} {\bibfnamefont {A.}~\bibnamefont {Aliper}}, \bibinfo {author} {\bibfnamefont {D.}~\bibnamefont {Polykovsky}}, \bibinfo {author} {\bibfnamefont {K.~M.~P.}\ \bibnamefont {Das}}, \bibinfo {author} {\bibfnamefont {J.}~\bibnamefont {Snider}}, \bibinfo {author} {\bibfnamefont {A.}~\bibnamefont {Lyakisheva}}, \bibinfo {author} {\bibfnamefont {A.~H.}\ \bibnamefont {Mansob}}, \bibinfo {author} {\bibfnamefont {Z.}~\bibnamefont {Yao}}, \bibinfo {author} {\bibfnamefont {L.}~\bibnamefont {Bitar}}, \bibinfo {author} {\bibfnamefont {E.}~\bibnamefont {Radchenko}}, \bibinfo {author} {\bibfnamefont {X.}~\bibnamefont {Ding}}, \bibinfo {author} {\bibfnamefont {J.}~\bibnamefont
  {Liu}}, \bibinfo {author} {\bibfnamefont {F.}~\bibnamefont {Meng}}, \bibinfo {author} {\bibfnamefont {F.}~\bibnamefont {Ren}}, \bibinfo {author} {\bibfnamefont {Y.}~\bibnamefont {Cao}}, \bibinfo {author} {\bibfnamefont {I.}~\bibnamefont {Stagljar}}, \bibinfo {author} {\bibfnamefont {A.}~\bibnamefont {Aspuru-Guzik}}, \ and\ \bibinfo {author} {\bibfnamefont {A.}~\bibnamefont {Zhavoronkov}},\ }\href {https://arxiv.org/abs/2402.08210} {\enquote {\bibinfo {title} {Quantum computing-enhanced algorithm unveils novel inhibitors for kras},}\ } (\bibinfo {year} {2024}),\ \Eprint {http://arxiv.org/abs/2402.08210} {arXiv:2402.08210 [quant-ph]} \BibitemShut {NoStop}%
\bibitem [{\citenamefont {Jaeger}\ and\ \citenamefont {Haas}(2004)}]{Jaeger2004}%
  \BibitemOpen
  \bibfield  {author} {\bibinfo {author} {\bibfnamefont {H.}~\bibnamefont {Jaeger}}\ and\ \bibinfo {author} {\bibfnamefont {H.}~\bibnamefont {Haas}},\ }\href {\doibase 10.1126/science.1091277} {\bibfield  {journal} {\bibinfo  {journal} {Science}\ }\textbf {\bibinfo {volume} {304}},\ \bibinfo {pages} {78} (\bibinfo {year} {2004})}\BibitemShut {NoStop}%
\bibitem [{\citenamefont {McClean}\ \emph {et~al.}(2018)\citenamefont {McClean}, \citenamefont {Boixo}, \citenamefont {Smelyanskiy}, \citenamefont {Babbush},\ and\ \citenamefont {Neven}}]{McClean_2018}%
  \BibitemOpen
  \bibfield  {author} {\bibinfo {author} {\bibfnamefont {J.~R.}\ \bibnamefont {McClean}}, \bibinfo {author} {\bibfnamefont {S.}~\bibnamefont {Boixo}}, \bibinfo {author} {\bibfnamefont {V.~N.}\ \bibnamefont {Smelyanskiy}}, \bibinfo {author} {\bibfnamefont {R.}~\bibnamefont {Babbush}}, \ and\ \bibinfo {author} {\bibfnamefont {H.}~\bibnamefont {Neven}},\ }\href {\doibase 10.1038/s41467-018-07090-4} {\bibfield  {journal} {\bibinfo  {journal} {Nature Communications}\ }\textbf {\bibinfo {volume} {9}} (\bibinfo {year} {2018}),\ 10.1038/s41467-018-07090-4}\BibitemShut {NoStop}%
\bibitem [{\citenamefont {Marrero}\ \emph {et~al.}(2021)\citenamefont {Marrero}, \citenamefont {Kieferová},\ and\ \citenamefont {Wiebe}}]{marrero2021entanglementinducedbarrenplateaus}%
  \BibitemOpen
  \bibfield  {author} {\bibinfo {author} {\bibfnamefont {C.~O.}\ \bibnamefont {Marrero}}, \bibinfo {author} {\bibfnamefont {M.}~\bibnamefont {Kieferová}}, \ and\ \bibinfo {author} {\bibfnamefont {N.}~\bibnamefont {Wiebe}},\ }\href {https://arxiv.org/abs/2010.15968} {\enquote {\bibinfo {title} {Entanglement induced barren plateaus},}\ } (\bibinfo {year} {2021}),\ \Eprint {http://arxiv.org/abs/2010.15968} {arXiv:2010.15968 [quant-ph]} \BibitemShut {NoStop}%
\bibitem [{\citenamefont {Larocca}\ \emph {et~al.}(2024)\citenamefont {Larocca}, \citenamefont {Thanasilp}, \citenamefont {Wang}, \citenamefont {Sharma}, \citenamefont {Biamonte}, \citenamefont {Coles}, \citenamefont {Cincio}, \citenamefont {McClean}, \citenamefont {Holmes},\ and\ \citenamefont {Cerezo}}]{larocca2024reviewbarrenplateausvariational}%
  \BibitemOpen
  \bibfield  {author} {\bibinfo {author} {\bibfnamefont {M.}~\bibnamefont {Larocca}}, \bibinfo {author} {\bibfnamefont {S.}~\bibnamefont {Thanasilp}}, \bibinfo {author} {\bibfnamefont {S.}~\bibnamefont {Wang}}, \bibinfo {author} {\bibfnamefont {K.}~\bibnamefont {Sharma}}, \bibinfo {author} {\bibfnamefont {J.}~\bibnamefont {Biamonte}}, \bibinfo {author} {\bibfnamefont {P.~J.}\ \bibnamefont {Coles}}, \bibinfo {author} {\bibfnamefont {L.}~\bibnamefont {Cincio}}, \bibinfo {author} {\bibfnamefont {J.~R.}\ \bibnamefont {McClean}}, \bibinfo {author} {\bibfnamefont {Z.}~\bibnamefont {Holmes}}, \ and\ \bibinfo {author} {\bibfnamefont {M.}~\bibnamefont {Cerezo}},\ }\href {https://arxiv.org/abs/2405.00781} {\enquote {\bibinfo {title} {A review of barren plateaus in variational quantum computing},}\ } (\bibinfo {year} {2024}),\ \Eprint {http://arxiv.org/abs/2405.00781} {arXiv:2405.00781 [quant-ph]} \BibitemShut {NoStop}%
\bibitem [{\citenamefont {Fujii}\ and\ \citenamefont {Nakajima}(2017)}]{Fujii_2017}%
  \BibitemOpen
  \bibfield  {author} {\bibinfo {author} {\bibfnamefont {K.}~\bibnamefont {Fujii}}\ and\ \bibinfo {author} {\bibfnamefont {K.}~\bibnamefont {Nakajima}},\ }\href {\doibase 10.1103/physrevapplied.8.024030} {\bibfield  {journal} {\bibinfo  {journal} {Physical Review Applied}\ }\textbf {\bibinfo {volume} {8}} (\bibinfo {year} {2017}),\ 10.1103/physrevapplied.8.024030}\BibitemShut {NoStop}%
\bibitem [{\citenamefont {Mart\'{\i}nez-Pe\~na}\ \emph {et~al.}(2021)\citenamefont {Mart\'{\i}nez-Pe\~na}, \citenamefont {Giorgi}, \citenamefont {Nokkala}, \citenamefont {Soriano},\ and\ \citenamefont {Zambrini}}]{Martinez2021}%
  \BibitemOpen
  \bibfield  {author} {\bibinfo {author} {\bibfnamefont {R.}~\bibnamefont {Mart\'{\i}nez-Pe\~na}}, \bibinfo {author} {\bibfnamefont {G.~L.}\ \bibnamefont {Giorgi}}, \bibinfo {author} {\bibfnamefont {J.}~\bibnamefont {Nokkala}}, \bibinfo {author} {\bibfnamefont {M.~C.}\ \bibnamefont {Soriano}}, \ and\ \bibinfo {author} {\bibfnamefont {R.}~\bibnamefont {Zambrini}},\ }\href {\doibase 10.1103/PhysRevLett.127.100502} {\bibfield  {journal} {\bibinfo  {journal} {Phys. Rev. Lett.}\ }\textbf {\bibinfo {volume} {127}},\ \bibinfo {pages} {100502} (\bibinfo {year} {2021})}\BibitemShut {NoStop}%
\bibitem [{\citenamefont {Bravo}\ \emph {et~al.}(2022)\citenamefont {Bravo}, \citenamefont {Najafi}, \citenamefont {Gao},\ and\ \citenamefont {Yelin}}]{Bravo2022}%
  \BibitemOpen
  \bibfield  {author} {\bibinfo {author} {\bibfnamefont {R.~A.}\ \bibnamefont {Bravo}}, \bibinfo {author} {\bibfnamefont {K.}~\bibnamefont {Najafi}}, \bibinfo {author} {\bibfnamefont {X.}~\bibnamefont {Gao}}, \ and\ \bibinfo {author} {\bibfnamefont {S.~F.}\ \bibnamefont {Yelin}},\ }\href {\doibase 10.1103/PRXQuantum.3.030325} {\bibfield  {journal} {\bibinfo  {journal} {PRX Quantum}\ }\textbf {\bibinfo {volume} {3}},\ \bibinfo {pages} {030325} (\bibinfo {year} {2022})}\BibitemShut {NoStop}%
\bibitem [{\citenamefont {Senanian}\ \emph {et~al.}(2023)\citenamefont {Senanian}, \citenamefont {Prabhu}, \citenamefont {Kremenetski}, \citenamefont {Roy}, \citenamefont {Cao}, \citenamefont {Kline}, \citenamefont {Onodera}, \citenamefont {Wright}, \citenamefont {Wu}, \citenamefont {Fatemi},\ and\ \citenamefont {McMahon}}]{Senanian2023}%
  \BibitemOpen
  \bibfield  {author} {\bibinfo {author} {\bibfnamefont {A.}~\bibnamefont {Senanian}}, \bibinfo {author} {\bibfnamefont {S.}~\bibnamefont {Prabhu}}, \bibinfo {author} {\bibfnamefont {V.}~\bibnamefont {Kremenetski}}, \bibinfo {author} {\bibfnamefont {S.}~\bibnamefont {Roy}}, \bibinfo {author} {\bibfnamefont {Y.}~\bibnamefont {Cao}}, \bibinfo {author} {\bibfnamefont {J.}~\bibnamefont {Kline}}, \bibinfo {author} {\bibfnamefont {T.}~\bibnamefont {Onodera}}, \bibinfo {author} {\bibfnamefont {L.~G.}\ \bibnamefont {Wright}}, \bibinfo {author} {\bibfnamefont {X.}~\bibnamefont {Wu}}, \bibinfo {author} {\bibfnamefont {V.}~\bibnamefont {Fatemi}}, \ and\ \bibinfo {author} {\bibfnamefont {P.~L.}\ \bibnamefont {McMahon}},\ }\href@noop {} {\enquote {\bibinfo {title} {Microwave signal processing using an analog quantum reservoir computer},}\ } (\bibinfo {year} {2023}),\ \Eprint {http://arxiv.org/abs/2312.16166} {arXiv:2312.16166 [quant-ph]} \BibitemShut {NoStop}%
\bibitem [{\citenamefont {Scholl}\ \emph {et~al.}(2021)\citenamefont {Scholl}, \citenamefont {Schuler}, \citenamefont {Williams}, \citenamefont {Eberharter}, \citenamefont {Barredo}, \citenamefont {Schymik}, \citenamefont {Lienhard}, \citenamefont {Henry}, \citenamefont {Lang}, \citenamefont {Lahaye}, \citenamefont {La\"uchli},\ and\ \citenamefont {Browaeys}}]{Scholl2021}%
  \BibitemOpen
  \bibfield  {author} {\bibinfo {author} {\bibfnamefont {P.}~\bibnamefont {Scholl}}, \bibinfo {author} {\bibfnamefont {M.}~\bibnamefont {Schuler}}, \bibinfo {author} {\bibfnamefont {H.~J.}\ \bibnamefont {Williams}}, \bibinfo {author} {\bibfnamefont {A.~A.}\ \bibnamefont {Eberharter}}, \bibinfo {author} {\bibfnamefont {D.}~\bibnamefont {Barredo}}, \bibinfo {author} {\bibfnamefont {K.-N.}\ \bibnamefont {Schymik}}, \bibinfo {author} {\bibfnamefont {V.}~\bibnamefont {Lienhard}}, \bibinfo {author} {\bibfnamefont {L.-P.}\ \bibnamefont {Henry}}, \bibinfo {author} {\bibfnamefont {T.~C.}\ \bibnamefont {Lang}}, \bibinfo {author} {\bibfnamefont {T.}~\bibnamefont {Lahaye}}, \bibinfo {author} {\bibfnamefont {A.~M.}\ \bibnamefont {La\"uchli}}, \ and\ \bibinfo {author} {\bibfnamefont {A.}~\bibnamefont {Browaeys}},\ }\href {\doibase https://doi.org/10.1038/s41586-021-03585-1} {\bibfield  {journal} {\bibinfo  {journal} {Nature}\ }\textbf {\bibinfo {volume} {595}},\ \bibinfo {pages} {233} (\bibinfo {year} {2021})}\BibitemShut
  {NoStop}%
\bibitem [{\citenamefont {Ebadi}\ \emph {et~al.}(2021)\citenamefont {Ebadi}, \citenamefont {Wang}, \citenamefont {Levine}, \citenamefont {Keesling}, \citenamefont {Semeghini}, \citenamefont {Omran}, \citenamefont {Bluvstein}, \citenamefont {Samajdar}, \citenamefont {Pichler}, \citenamefont {Ho}, \citenamefont {Choi}, \citenamefont {Sachdev}, \citenamefont {Greiner}, \citenamefont {Vuleti{\'{c}}},\ and\ \citenamefont {Lukin}}]{Ebadi2021}%
  \BibitemOpen
  \bibfield  {author} {\bibinfo {author} {\bibfnamefont {S.}~\bibnamefont {Ebadi}}, \bibinfo {author} {\bibfnamefont {T.~T.}\ \bibnamefont {Wang}}, \bibinfo {author} {\bibfnamefont {H.}~\bibnamefont {Levine}}, \bibinfo {author} {\bibfnamefont {A.}~\bibnamefont {Keesling}}, \bibinfo {author} {\bibfnamefont {G.}~\bibnamefont {Semeghini}}, \bibinfo {author} {\bibfnamefont {A.}~\bibnamefont {Omran}}, \bibinfo {author} {\bibfnamefont {D.}~\bibnamefont {Bluvstein}}, \bibinfo {author} {\bibfnamefont {R.}~\bibnamefont {Samajdar}}, \bibinfo {author} {\bibfnamefont {H.}~\bibnamefont {Pichler}}, \bibinfo {author} {\bibfnamefont {W.~W.}\ \bibnamefont {Ho}}, \bibinfo {author} {\bibfnamefont {S.}~\bibnamefont {Choi}}, \bibinfo {author} {\bibfnamefont {S.}~\bibnamefont {Sachdev}}, \bibinfo {author} {\bibfnamefont {M.}~\bibnamefont {Greiner}}, \bibinfo {author} {\bibfnamefont {V.}~\bibnamefont {Vuleti{\'{c}}}}, \ and\ \bibinfo {author} {\bibfnamefont {M.~D.}\ \bibnamefont {Lukin}},\ }\href {\doibase
  10.1038/s41586-021-03582-4} {\bibfield  {journal} {\bibinfo  {journal} {Nature}\ }\textbf {\bibinfo {volume} {595}},\ \bibinfo {pages} {227} (\bibinfo {year} {2021})}\BibitemShut {NoStop}%
\bibitem [{\citenamefont {Semeghini}\ \emph {et~al.}(2021)\citenamefont {Semeghini}, \citenamefont {Levine}, \citenamefont {Keesling}, \citenamefont {Ebadi}, \citenamefont {Wang}, \citenamefont {Bluvstein}, \citenamefont {Verresen}, \citenamefont {Pichler}, \citenamefont {Kalinowski}, \citenamefont {Samajdar}, \citenamefont {Omran}, \citenamefont {Sachdev}, \citenamefont {Vishwanath}, \citenamefont {Greiner}, \citenamefont {Vuletić},\ and\ \citenamefont {Lukin}}]{Semeghini2021}%
  \BibitemOpen
  \bibfield  {author} {\bibinfo {author} {\bibfnamefont {G.}~\bibnamefont {Semeghini}}, \bibinfo {author} {\bibfnamefont {H.}~\bibnamefont {Levine}}, \bibinfo {author} {\bibfnamefont {A.}~\bibnamefont {Keesling}}, \bibinfo {author} {\bibfnamefont {S.}~\bibnamefont {Ebadi}}, \bibinfo {author} {\bibfnamefont {T.~T.}\ \bibnamefont {Wang}}, \bibinfo {author} {\bibfnamefont {D.}~\bibnamefont {Bluvstein}}, \bibinfo {author} {\bibfnamefont {R.}~\bibnamefont {Verresen}}, \bibinfo {author} {\bibfnamefont {H.}~\bibnamefont {Pichler}}, \bibinfo {author} {\bibfnamefont {M.}~\bibnamefont {Kalinowski}}, \bibinfo {author} {\bibfnamefont {R.}~\bibnamefont {Samajdar}}, \bibinfo {author} {\bibfnamefont {A.}~\bibnamefont {Omran}}, \bibinfo {author} {\bibfnamefont {S.}~\bibnamefont {Sachdev}}, \bibinfo {author} {\bibfnamefont {A.}~\bibnamefont {Vishwanath}}, \bibinfo {author} {\bibfnamefont {M.}~\bibnamefont {Greiner}}, \bibinfo {author} {\bibfnamefont {V.}~\bibnamefont {Vuletić}}, \ and\ \bibinfo {author} {\bibfnamefont
  {M.~D.}\ \bibnamefont {Lukin}},\ }\href {\doibase 10.1126/science.abi8794} {\bibfield  {journal} {\bibinfo  {journal} {Science}\ }\textbf {\bibinfo {volume} {374}},\ \bibinfo {pages} {1242} (\bibinfo {year} {2021})}\BibitemShut {NoStop}%
\bibitem [{\citenamefont {Ebadi}\ \emph {et~al.}(2022)\citenamefont {Ebadi}, \citenamefont {Keesling}, \citenamefont {Cain}, \citenamefont {Wang}, \citenamefont {Levine}, \citenamefont {Bluvstein}, \citenamefont {Semeghini}, \citenamefont {Omran}, \citenamefont {Liu}, \citenamefont {Samajdar}, \citenamefont {Luo}, \citenamefont {Nash}, \citenamefont {Gao}, \citenamefont {Barak}, \citenamefont {Farhi}, \citenamefont {Sachdev}, \citenamefont {Gemelke}, \citenamefont {Zhou}, \citenamefont {Choi}, \citenamefont {Pichler}, \citenamefont {Wang}, \citenamefont {Greiner}, \citenamefont {Vuletić},\ and\ \citenamefont {Lukin}}]{Ebadi2022}%
  \BibitemOpen
  \bibfield  {author} {\bibinfo {author} {\bibfnamefont {S.}~\bibnamefont {Ebadi}}, \bibinfo {author} {\bibfnamefont {A.}~\bibnamefont {Keesling}}, \bibinfo {author} {\bibfnamefont {M.}~\bibnamefont {Cain}}, \bibinfo {author} {\bibfnamefont {T.~T.}\ \bibnamefont {Wang}}, \bibinfo {author} {\bibfnamefont {H.}~\bibnamefont {Levine}}, \bibinfo {author} {\bibfnamefont {D.}~\bibnamefont {Bluvstein}}, \bibinfo {author} {\bibfnamefont {G.}~\bibnamefont {Semeghini}}, \bibinfo {author} {\bibfnamefont {A.}~\bibnamefont {Omran}}, \bibinfo {author} {\bibfnamefont {J.-G.}\ \bibnamefont {Liu}}, \bibinfo {author} {\bibfnamefont {R.}~\bibnamefont {Samajdar}}, \bibinfo {author} {\bibfnamefont {X.-Z.}\ \bibnamefont {Luo}}, \bibinfo {author} {\bibfnamefont {B.}~\bibnamefont {Nash}}, \bibinfo {author} {\bibfnamefont {X.}~\bibnamefont {Gao}}, \bibinfo {author} {\bibfnamefont {B.}~\bibnamefont {Barak}}, \bibinfo {author} {\bibfnamefont {E.}~\bibnamefont {Farhi}}, \bibinfo {author} {\bibfnamefont {S.}~\bibnamefont {Sachdev}},
  \bibinfo {author} {\bibfnamefont {N.}~\bibnamefont {Gemelke}}, \bibinfo {author} {\bibfnamefont {L.}~\bibnamefont {Zhou}}, \bibinfo {author} {\bibfnamefont {S.}~\bibnamefont {Choi}}, \bibinfo {author} {\bibfnamefont {H.}~\bibnamefont {Pichler}}, \bibinfo {author} {\bibfnamefont {S.-T.}\ \bibnamefont {Wang}}, \bibinfo {author} {\bibfnamefont {M.}~\bibnamefont {Greiner}}, \bibinfo {author} {\bibfnamefont {V.}~\bibnamefont {Vuletić}}, \ and\ \bibinfo {author} {\bibfnamefont {M.~D.}\ \bibnamefont {Lukin}},\ }\href {\doibase 10.1126/science.abo6587} {\bibfield  {journal} {\bibinfo  {journal} {Science}\ }\textbf {\bibinfo {volume} {376}},\ \bibinfo {pages} {1209} (\bibinfo {year} {2022})}\BibitemShut {NoStop}%
\bibitem [{\citenamefont {Wurtz}\ \emph {et~al.}(2023)\citenamefont {Wurtz}, \citenamefont {Bylinskii}, \citenamefont {Braverman}, \citenamefont {Amato-Grill}, \citenamefont {Cantu}, \citenamefont {Huber}, \citenamefont {Lukin}, \citenamefont {Liu}, \citenamefont {Weinberg}, \citenamefont {Long}, \citenamefont {Wang}, \citenamefont {Gemelke},\ and\ \citenamefont {Keesling}}]{wurtz2023aquila}%
  \BibitemOpen
  \bibfield  {author} {\bibinfo {author} {\bibfnamefont {J.}~\bibnamefont {Wurtz}}, \bibinfo {author} {\bibfnamefont {A.}~\bibnamefont {Bylinskii}}, \bibinfo {author} {\bibfnamefont {B.}~\bibnamefont {Braverman}}, \bibinfo {author} {\bibfnamefont {J.}~\bibnamefont {Amato-Grill}}, \bibinfo {author} {\bibfnamefont {S.~H.}\ \bibnamefont {Cantu}}, \bibinfo {author} {\bibfnamefont {F.}~\bibnamefont {Huber}}, \bibinfo {author} {\bibfnamefont {A.}~\bibnamefont {Lukin}}, \bibinfo {author} {\bibfnamefont {F.}~\bibnamefont {Liu}}, \bibinfo {author} {\bibfnamefont {P.}~\bibnamefont {Weinberg}}, \bibinfo {author} {\bibfnamefont {J.}~\bibnamefont {Long}}, \bibinfo {author} {\bibfnamefont {S.-T.}\ \bibnamefont {Wang}}, \bibinfo {author} {\bibfnamefont {N.}~\bibnamefont {Gemelke}}, \ and\ \bibinfo {author} {\bibfnamefont {A.}~\bibnamefont {Keesling}},\ }\href@noop {} {\enquote {\bibinfo {title} {Aquila: Quera's 256-qubit neutral-atom quantum computer},}\ } (\bibinfo {year} {2023}),\ \Eprint {http://arxiv.org/abs/2306.11727}
  {arXiv:2306.11727 [quant-ph]} \BibitemShut {NoStop}%
\bibitem [{\citenamefont {Balewski}\ \emph {et~al.}(2024)\citenamefont {Balewski}, \citenamefont {Kornjaca}, \citenamefont {Klymko}, \citenamefont {Darbha}, \citenamefont {Hirsbrunner}, \citenamefont {Lopes}, \citenamefont {Liu},\ and\ \citenamefont {Camps}}]{quera2024}%
  \BibitemOpen
  \bibfield  {author} {\bibinfo {author} {\bibfnamefont {J.}~\bibnamefont {Balewski}}, \bibinfo {author} {\bibfnamefont {M.}~\bibnamefont {Kornjaca}}, \bibinfo {author} {\bibfnamefont {K.}~\bibnamefont {Klymko}}, \bibinfo {author} {\bibfnamefont {S.}~\bibnamefont {Darbha}}, \bibinfo {author} {\bibfnamefont {M.~R.}\ \bibnamefont {Hirsbrunner}}, \bibinfo {author} {\bibfnamefont {P.}~\bibnamefont {Lopes}}, \bibinfo {author} {\bibfnamefont {F.}~\bibnamefont {Liu}}, \ and\ \bibinfo {author} {\bibfnamefont {D.}~\bibnamefont {Camps}},\ }\href@noop {} {\enquote {\bibinfo {title} {Engineering quantum states with neutral atoms},}\ } (\bibinfo {year} {2024}),\ \Eprint {http://arxiv.org/abs/2404.04411} {arXiv:2404.04411 [quant-ph]} \BibitemShut {NoStop}%
\bibitem [{\citenamefont {Bluvstein}\ \emph {et~al.}(2023)\citenamefont {Bluvstein}, \citenamefont {Evered}, \citenamefont {Geim}, \citenamefont {Li}, \citenamefont {Zhou}, \citenamefont {Manovitz}, \citenamefont {Ebadi}, \citenamefont {Cain}, \citenamefont {Kalinowski}, \citenamefont {Hangleiter}, \citenamefont {Bonilla~Ataides}, \citenamefont {Maskara}, \citenamefont {Cong}, \citenamefont {Gao}, \citenamefont {Sales~Rodriguez}, \citenamefont {Karolyshyn}, \citenamefont {Semeghini}, \citenamefont {Gullans}, \citenamefont {Greiner}, \citenamefont {Vuletić},\ and\ \citenamefont {Lukin}}]{Bluvstein2023}%
  \BibitemOpen
  \bibfield  {author} {\bibinfo {author} {\bibfnamefont {D.}~\bibnamefont {Bluvstein}}, \bibinfo {author} {\bibfnamefont {S.~J.}\ \bibnamefont {Evered}}, \bibinfo {author} {\bibfnamefont {A.~A.}\ \bibnamefont {Geim}}, \bibinfo {author} {\bibfnamefont {S.~H.}\ \bibnamefont {Li}}, \bibinfo {author} {\bibfnamefont {H.}~\bibnamefont {Zhou}}, \bibinfo {author} {\bibfnamefont {T.}~\bibnamefont {Manovitz}}, \bibinfo {author} {\bibfnamefont {S.}~\bibnamefont {Ebadi}}, \bibinfo {author} {\bibfnamefont {M.}~\bibnamefont {Cain}}, \bibinfo {author} {\bibfnamefont {M.}~\bibnamefont {Kalinowski}}, \bibinfo {author} {\bibfnamefont {D.}~\bibnamefont {Hangleiter}}, \bibinfo {author} {\bibfnamefont {J.~P.}\ \bibnamefont {Bonilla~Ataides}}, \bibinfo {author} {\bibfnamefont {N.}~\bibnamefont {Maskara}}, \bibinfo {author} {\bibfnamefont {I.}~\bibnamefont {Cong}}, \bibinfo {author} {\bibfnamefont {X.}~\bibnamefont {Gao}}, \bibinfo {author} {\bibfnamefont {P.}~\bibnamefont {Sales~Rodriguez}}, \bibinfo {author} {\bibfnamefont
  {T.}~\bibnamefont {Karolyshyn}}, \bibinfo {author} {\bibfnamefont {G.}~\bibnamefont {Semeghini}}, \bibinfo {author} {\bibfnamefont {M.~J.}\ \bibnamefont {Gullans}}, \bibinfo {author} {\bibfnamefont {M.}~\bibnamefont {Greiner}}, \bibinfo {author} {\bibfnamefont {V.}~\bibnamefont {Vuletić}}, \ and\ \bibinfo {author} {\bibfnamefont {M.~D.}\ \bibnamefont {Lukin}},\ }\href {\doibase 10.1038/s41586-023-06927-3} {\bibfield  {journal} {\bibinfo  {journal} {Nature}\ }\textbf {\bibinfo {volume} {626}},\ \bibinfo {pages} {58–65} (\bibinfo {year} {2023})}\BibitemShut {NoStop}%
\bibitem [{\citenamefont {Kulkarni}(2012)}]{merck2012}%
  \BibitemOpen
  \bibfield  {author} {\bibinfo {author} {\bibfnamefont {A.}~\bibnamefont {Kulkarni}},\ }\href@noop {} {\enquote {\bibinfo {title} {Merck molecular activity challenge},}\ }\bibinfo {howpublished} {\url{https://kaggle.com/competitions/MerckActivity}} (\bibinfo {year} {2012}),\ \bibinfo {note} {accessed: 2024-06-04}\BibitemShut {NoStop}%
\bibitem [{Note1()}]{Note1}%
  \BibitemOpen
  \bibinfo {note} {The Competition was sponsored in 2012 by Merck \& Co., Inc, located today at 26 E Lincoln Ave, Rahway, NJ 07065, US, USA. Merck KGaA and Merck \& Co.~are separate entities. Merck KGaA is headquartered in Germany, operates globally as EMD in North America, focusing on pharmaceuticals, life sciences, and performance materials, whereas Merck \& Co., headquartered in the USA, operates as MSD outside North America, primarily focusing on pharmaceuticals, vaccines, and animal health.}\BibitemShut {Stop}%
\bibitem [{\citenamefont {Ramsundar}\ \emph {et~al.}(2015)\citenamefont {Ramsundar}, \citenamefont {Kearnes}, \citenamefont {Riley}, \citenamefont {Webster}, \citenamefont {Konerding},\ and\ \citenamefont {Pande}}]{ramsundar2015massivelymultitasknetworksdrug}%
  \BibitemOpen
  \bibfield  {author} {\bibinfo {author} {\bibfnamefont {B.}~\bibnamefont {Ramsundar}}, \bibinfo {author} {\bibfnamefont {S.}~\bibnamefont {Kearnes}}, \bibinfo {author} {\bibfnamefont {P.}~\bibnamefont {Riley}}, \bibinfo {author} {\bibfnamefont {D.}~\bibnamefont {Webster}}, \bibinfo {author} {\bibfnamefont {D.}~\bibnamefont {Konerding}}, \ and\ \bibinfo {author} {\bibfnamefont {V.}~\bibnamefont {Pande}},\ }\href {https://arxiv.org/abs/1502.02072} {\enquote {\bibinfo {title} {Massively multitask networks for drug discovery},}\ } (\bibinfo {year} {2015}),\ \Eprint {http://arxiv.org/abs/1502.02072} {arXiv:1502.02072 [stat.ML]} \BibitemShut {NoStop}%
\bibitem [{\citenamefont {Goh}\ \emph {et~al.}(2017)\citenamefont {Goh}, \citenamefont {Siegel}, \citenamefont {Vishnu}, \citenamefont {Hodas},\ and\ \citenamefont {Baker}}]{goh2017chemceptiondeepneuralnetwork}%
  \BibitemOpen
  \bibfield  {author} {\bibinfo {author} {\bibfnamefont {G.~B.}\ \bibnamefont {Goh}}, \bibinfo {author} {\bibfnamefont {C.}~\bibnamefont {Siegel}}, \bibinfo {author} {\bibfnamefont {A.}~\bibnamefont {Vishnu}}, \bibinfo {author} {\bibfnamefont {N.~O.}\ \bibnamefont {Hodas}}, \ and\ \bibinfo {author} {\bibfnamefont {N.}~\bibnamefont {Baker}},\ }\href {https://arxiv.org/abs/1706.06689} {\enquote {\bibinfo {title} {Chemception: A deep neural network with minimal chemistry knowledge matches the performance of expert-developed qsar/qspr models},}\ } (\bibinfo {year} {2017}),\ \Eprint {http://arxiv.org/abs/1706.06689} {arXiv:1706.06689 [stat.ML]} \BibitemShut {NoStop}%
\bibitem [{\citenamefont {Hechtlinger}\ \emph {et~al.}(2017)\citenamefont {Hechtlinger}, \citenamefont {Chakravarti},\ and\ \citenamefont {Qin}}]{hechtlinger2017generalizationconvolutionalneuralnetworks}%
  \BibitemOpen
  \bibfield  {author} {\bibinfo {author} {\bibfnamefont {Y.}~\bibnamefont {Hechtlinger}}, \bibinfo {author} {\bibfnamefont {P.}~\bibnamefont {Chakravarti}}, \ and\ \bibinfo {author} {\bibfnamefont {J.}~\bibnamefont {Qin}},\ }\href {https://arxiv.org/abs/1704.08165} {\enquote {\bibinfo {title} {A generalization of convolutional neural networks to graph-structured data},}\ } (\bibinfo {year} {2017}),\ \Eprint {http://arxiv.org/abs/1704.08165} {arXiv:1704.08165 [stat.ML]} \BibitemShut {NoStop}%
\bibitem [{\citenamefont {McInnes}\ \emph {et~al.}(2020)\citenamefont {McInnes}, \citenamefont {Healy},\ and\ \citenamefont {Melville}}]{mcinnes2020umapuniformmanifoldapproximation}%
  \BibitemOpen
  \bibfield  {author} {\bibinfo {author} {\bibfnamefont {L.}~\bibnamefont {McInnes}}, \bibinfo {author} {\bibfnamefont {J.}~\bibnamefont {Healy}}, \ and\ \bibinfo {author} {\bibfnamefont {J.}~\bibnamefont {Melville}},\ }\href {https://arxiv.org/abs/1802.03426} {\enquote {\bibinfo {title} {Umap: Uniform manifold approximation and projection for dimension reduction},}\ } (\bibinfo {year} {2020}),\ \Eprint {http://arxiv.org/abs/1802.03426} {arXiv:1802.03426 [stat.ML]} \BibitemShut {NoStop}%
\bibitem [{\citenamefont {Lundberg}\ and\ \citenamefont {Lee}(2017)}]{10.5555/3295222.3295230}%
  \BibitemOpen
  \bibfield  {author} {\bibinfo {author} {\bibfnamefont {S.~M.}\ \bibnamefont {Lundberg}}\ and\ \bibinfo {author} {\bibfnamefont {S.-I.}\ \bibnamefont {Lee}},\ }in\ \href@noop {} {\emph {\bibinfo {booktitle} {Proceedings of the 31st International Conference on Neural Information Processing Systems}}},\ \bibinfo {series and number} {NIPS'17}\ (\bibinfo  {publisher} {Curran Associates Inc.},\ \bibinfo {address} {Red Hook, NY, USA},\ \bibinfo {year} {2017})\ p.\ \bibinfo {pages} {4768–4777}\BibitemShut {NoStop}%
\bibitem [{\citenamefont {Team}({\natexlab{b}})}]{shapintro}%
  \BibitemOpen
  \bibfield  {author} {\bibinfo {author} {\bibfnamefont {S.}~\bibnamefont {Team}},\ }\href {https://shap.readthedocs.io/en/latest/example_notebooks/overviews/An%20introduction%20to%20explainable%20AI%20with%20Shapley%20values.html} {\enquote {\bibinfo {title} {Introduction to shap},}\ } ({\natexlab{b}}),\ \bibinfo {note} {accessed: 2024-09-13}\BibitemShut {NoStop}%
\bibitem [{\citenamefont {Pedregosa}\ \emph {et~al.}(2011)\citenamefont {Pedregosa}, \citenamefont {Varoquaux}, \citenamefont {Gramfort}, \citenamefont {Michel}, \citenamefont {Thirion}, \citenamefont {Grisel}, \citenamefont {Blondel}, \citenamefont {Prettenhofer}, \citenamefont {Weiss}, \citenamefont {Dubourg}, \citenamefont {Vanderplas}, \citenamefont {Passos}, \citenamefont {Cournapeau}, \citenamefont {Brucher}, \citenamefont {Perrot},\ and\ \citenamefont {Duchesnay}}]{scikit-learn}%
  \BibitemOpen
  \bibfield  {author} {\bibinfo {author} {\bibfnamefont {F.}~\bibnamefont {Pedregosa}}, \bibinfo {author} {\bibfnamefont {G.}~\bibnamefont {Varoquaux}}, \bibinfo {author} {\bibfnamefont {A.}~\bibnamefont {Gramfort}}, \bibinfo {author} {\bibfnamefont {V.}~\bibnamefont {Michel}}, \bibinfo {author} {\bibfnamefont {B.}~\bibnamefont {Thirion}}, \bibinfo {author} {\bibfnamefont {O.}~\bibnamefont {Grisel}}, \bibinfo {author} {\bibfnamefont {M.}~\bibnamefont {Blondel}}, \bibinfo {author} {\bibfnamefont {P.}~\bibnamefont {Prettenhofer}}, \bibinfo {author} {\bibfnamefont {R.}~\bibnamefont {Weiss}}, \bibinfo {author} {\bibfnamefont {V.}~\bibnamefont {Dubourg}}, \bibinfo {author} {\bibfnamefont {J.}~\bibnamefont {Vanderplas}}, \bibinfo {author} {\bibfnamefont {A.}~\bibnamefont {Passos}}, \bibinfo {author} {\bibfnamefont {D.}~\bibnamefont {Cournapeau}}, \bibinfo {author} {\bibfnamefont {M.}~\bibnamefont {Brucher}}, \bibinfo {author} {\bibfnamefont {M.}~\bibnamefont {Perrot}}, \ and\ \bibinfo {author} {\bibfnamefont
  {E.}~\bibnamefont {Duchesnay}},\ }\href@noop {} {\bibfield  {journal} {\bibinfo  {journal} {Journal of Machine Learning Research}\ }\textbf {\bibinfo {volume} {12}},\ \bibinfo {pages} {2825} (\bibinfo {year} {2011})}\BibitemShut {NoStop}%
\bibitem [{\citenamefont {Team}({\natexlab{c}})}]{randomforestreg}%
  \BibitemOpen
  \bibfield  {author} {\bibinfo {author} {\bibfnamefont {S.~L.}\ \bibnamefont {Team}},\ }\href@noop {} {\enquote {\bibinfo {title} {Random forest regressor},}\ }\bibinfo {howpublished} {\url{https://scikit-learn.org/stable/modules/generated/sklearn.ensemble.RandomForestRegressor.html}} ({\natexlab{c}}),\ \bibinfo {note} {accessed: 2024-09-13}\BibitemShut {NoStop}%
\bibitem [{\citenamefont {Team}({\natexlab{d}})}]{gaussianprocessregressor}%
  \BibitemOpen
  \bibfield  {author} {\bibinfo {author} {\bibfnamefont {S.~L.}\ \bibnamefont {Team}},\ }\href@noop {} {\enquote {\bibinfo {title} {Forecasting of co2 level on mona loa dataset using gaussian process regression (gpr)},}\ }\bibinfo {howpublished} {\url{https://scikit-learn.org/stable/auto_examples/gaussian_process/plot_gpr_co2.html}} ({\natexlab{d}}),\ \bibinfo {note} {accessed: 2024-09-13}\BibitemShut {NoStop}%
\bibitem [{blo(2023)}]{bloqade2023quera}%
  \BibitemOpen
  \href {https://github.com/QuEraComputing/Bloqade.jl/} {\enquote {\bibinfo {title} {Bloqade.jl: {P}ackage for the quantum computation and quantum simulation based on the neutral-atom architecture.}}\ } (\bibinfo {year} {2023})\BibitemShut {NoStop}%
\bibitem [{QRC(2024)}]{QRCtutorials}%
  \BibitemOpen
  \href@noop {} {\enquote {\bibinfo {title} {{G}it{H}ub - {Q}u{E}ra{C}omputing/{Q}{R}{C}-tutorials: {A} set of tutorials for quantum reservoir learning. --- github.com},}\ }\bibinfo {howpublished} {\url{https://github.com/QuEraComputing/QRC-tutorials}} (\bibinfo {year} {2024}),\ \bibinfo {note} {[Accessed 04-10-2024]}\BibitemShut {NoStop}%
\bibitem [{\citenamefont {Dorrity}\ \emph {et~al.}(2020)\citenamefont {Dorrity}, \citenamefont {Saunders}, \citenamefont {Queitsch}, \citenamefont {Fields},\ and\ \citenamefont {Trapnell}}]{Dorrity2020}%
  \BibitemOpen
  \bibfield  {author} {\bibinfo {author} {\bibfnamefont {M.~W.}\ \bibnamefont {Dorrity}}, \bibinfo {author} {\bibfnamefont {L.~M.}\ \bibnamefont {Saunders}}, \bibinfo {author} {\bibfnamefont {C.}~\bibnamefont {Queitsch}}, \bibinfo {author} {\bibfnamefont {S.}~\bibnamefont {Fields}}, \ and\ \bibinfo {author} {\bibfnamefont {C.}~\bibnamefont {Trapnell}},\ }\href {\doibase 10.1038/s41467-020-15351-4} {\bibfield  {journal} {\bibinfo  {journal} {Nature Communications}\ }\textbf {\bibinfo {volume} {11}},\ \bibinfo {pages} {1537} (\bibinfo {year} {2020})}\BibitemShut {NoStop}%
\bibitem [{Note2()}]{Note2}%
  \BibitemOpen
  \bibinfo {note} {800 record samples for MMACD 5, 9, 14, and 15 features were chosen based on SHAP gradient method as compared to MMACD 4 which used SHAP kernel method to save on compute resources, as the change is verified to have minimal impact on the results.}\BibitemShut {Stop}%
\bibitem [{\citenamefont {Sachs}\ \emph {et~al.}(2005)\citenamefont {Sachs}, \citenamefont {Perez}, \citenamefont {Pe’er}, \citenamefont {Lauffenburger},\ and\ \citenamefont {Nolan}}]{Sachs_Perez_Pe’er_Lauffenburger_Nolan_2005}%
  \BibitemOpen
  \bibfield  {author} {\bibinfo {author} {\bibfnamefont {K.}~\bibnamefont {Sachs}}, \bibinfo {author} {\bibfnamefont {O.}~\bibnamefont {Perez}}, \bibinfo {author} {\bibfnamefont {D.}~\bibnamefont {Pe’er}}, \bibinfo {author} {\bibfnamefont {D.~A.}\ \bibnamefont {Lauffenburger}}, \ and\ \bibinfo {author} {\bibfnamefont {G.~P.}\ \bibnamefont {Nolan}},\ }\href {\doibase 10.1126/science.1105809} {\bibfield  {journal} {\bibinfo  {journal} {Science (New York, N.Y.)}\ }\textbf {\bibinfo {volume} {308}},\ \bibinfo {pages} {523–529} (\bibinfo {year} {2005})}\BibitemShut {NoStop}%
\bibitem [{\citenamefont {Gao}\ \emph {et~al.}(2022)\citenamefont {Gao}, \citenamefont {Anschuetz}, \citenamefont {Wang}, \citenamefont {Cirac},\ and\ \citenamefont {Lukin}}]{Gao2022}%
  \BibitemOpen
  \bibfield  {author} {\bibinfo {author} {\bibfnamefont {X.}~\bibnamefont {Gao}}, \bibinfo {author} {\bibfnamefont {E.~R.}\ \bibnamefont {Anschuetz}}, \bibinfo {author} {\bibfnamefont {S.-T.}\ \bibnamefont {Wang}}, \bibinfo {author} {\bibfnamefont {J.~I.}\ \bibnamefont {Cirac}}, \ and\ \bibinfo {author} {\bibfnamefont {M.~D.}\ \bibnamefont {Lukin}},\ }\href {\doibase 10.1103/physrevx.12.021037} {\bibfield  {journal} {\bibinfo  {journal} {Physical Review X}\ }\textbf {\bibinfo {volume} {12}} (\bibinfo {year} {2022}),\ 10.1103/physrevx.12.021037}\BibitemShut {NoStop}%
\bibitem [{\citenamefont {Team}({\natexlab{e}})}]{decisiontree}%
  \BibitemOpen
  \bibfield  {author} {\bibinfo {author} {\bibfnamefont {S.~L.}\ \bibnamefont {Team}},\ }\href@noop {} {\enquote {\bibinfo {title} {Decision trees},}\ }\bibinfo {howpublished} {\url{https://scikit-learn.org/stable/modules/tree.html}} ({\natexlab{e}}),\ \bibinfo {note} {accessed: 2024-09-13}\BibitemShut {NoStop}%
\bibitem [{\citenamefont {Team}({\natexlab{f}})}]{bagreg}%
  \BibitemOpen
  \bibfield  {author} {\bibinfo {author} {\bibfnamefont {S.~L.}\ \bibnamefont {Team}},\ }\href@noop {} {\enquote {\bibinfo {title} {Bagging regressor},}\ }\bibinfo {howpublished} {\url{https://scikit-learn.org/stable/modules/generated/sklearn.ensemble.BaggingRegressor.html}} ({\natexlab{f}}),\ \bibinfo {note} {accessed: 2024-09-13}\BibitemShut {NoStop}%
\bibitem [{\citenamefont {Team}({\natexlab{g}})}]{gradboost}%
  \BibitemOpen
  \bibfield  {author} {\bibinfo {author} {\bibfnamefont {S.~L.}\ \bibnamefont {Team}},\ }\href@noop {} {\enquote {\bibinfo {title} {Gradient boosting regression},}\ }\bibinfo {howpublished} {\url{https://scikit-learn.org/stable/auto_examples/ensemble/plot_gradient_boosting_regression.html}} ({\natexlab{g}}),\ \bibinfo {note} {accessed: 2024-09-13}\BibitemShut {NoStop}%
\bibitem [{\citenamefont {Team}({\natexlab{h}})}]{adaboost}%
  \BibitemOpen
  \bibfield  {author} {\bibinfo {author} {\bibfnamefont {S.~L.}\ \bibnamefont {Team}},\ }\href@noop {} {\enquote {\bibinfo {title} {Decision tree regression with adaboos},}\ }\bibinfo {howpublished} {\url{https://scikit-learn.org/stable/auto_examples/ensemble/plot_adaboost_regression.html}} ({\natexlab{h}}),\ \bibinfo {note} {accessed: 2024-09-13}\BibitemShut {NoStop}%
\bibitem [{\citenamefont {Team}({\natexlab{i}})}]{linreg}%
  \BibitemOpen
  \bibfield  {author} {\bibinfo {author} {\bibfnamefont {S.~L.}\ \bibnamefont {Team}},\ }\href@noop {} {\enquote {\bibinfo {title} {Linear regression},}\ }\bibinfo {howpublished} {\url{https://scikit-learn.org/stable/modules/linear_model.html}} ({\natexlab{i}}),\ \bibinfo {note} {accessed: 2024-09-13}\BibitemShut {NoStop}%
\bibitem [{\citenamefont {Team}({\natexlab{j}})}]{svr}%
  \BibitemOpen
  \bibfield  {author} {\bibinfo {author} {\bibfnamefont {S.~L.}\ \bibnamefont {Team}},\ }\href@noop {} {\enquote {\bibinfo {title} {Support vector regressor},}\ }\bibinfo {howpublished} {\url{https://scikit-learn.org/stable/modules/svm.html}} ({\natexlab{j}}),\ \bibinfo {note} {accessed: 2024-09-13}\BibitemShut {NoStop}%
\bibitem [{\citenamefont {Team}({\natexlab{k}})}]{kneighbor}%
  \BibitemOpen
  \bibfield  {author} {\bibinfo {author} {\bibfnamefont {S.~L.}\ \bibnamefont {Team}},\ }\href@noop {} {\enquote {\bibinfo {title} {Knearest neighbor},}\ }\bibinfo {howpublished} {\url{https://scikit-learn.org/stable/modules/neighbors.html}} ({\natexlab{k}}),\ \bibinfo {note} {accessed: 2024-09-13}\BibitemShut {NoStop}%
\bibitem [{\citenamefont {Team}({\natexlab{l}})}]{exttree}%
  \BibitemOpen
  \bibfield  {author} {\bibinfo {author} {\bibfnamefont {S.~L.}\ \bibnamefont {Team}},\ }\href@noop {} {\enquote {\bibinfo {title} {Extra tree regressor},}\ }\bibinfo {howpublished} {\url{https://scikit-learn.org/stable/modules/generated/sklearn.tree.ExtraTreeRegressor.html}} ({\natexlab{l}}),\ \bibinfo {note} {accessed: 2024-09-13}\BibitemShut {NoStop}%
\bibitem [{\citenamefont {Team}({\natexlab{m}})}]{sgdr}%
  \BibitemOpen
  \bibfield  {author} {\bibinfo {author} {\bibfnamefont {S.~L.}\ \bibnamefont {Team}},\ }\href@noop {} {\enquote {\bibinfo {title} {Sgd regressor},}\ }\bibinfo {howpublished} {\url{https://scikit-learn.org/stable/modules/generated/sklearn.linear_model.SGDRegressor.html}} ({\natexlab{m}}),\ \bibinfo {note} {accessed: 2024-09-13}\BibitemShut {NoStop}%
\bibitem [{\citenamefont {Schmidhuber}(2015)}]{Schmidhuber_2015}%
  \BibitemOpen
  \bibfield  {author} {\bibinfo {author} {\bibfnamefont {J.}~\bibnamefont {Schmidhuber}},\ }\href {\doibase 10.1016/j.neunet.2014.09.003} {\bibfield  {journal} {\bibinfo  {journal} {Neural Networks}\ }\textbf {\bibinfo {volume} {61}},\ \bibinfo {pages} {85–117} (\bibinfo {year} {2015})}\BibitemShut {NoStop}%
\bibitem [{\citenamefont {Chalise}\ \emph {et~al.}(2014)\citenamefont {Chalise}, \citenamefont {Koestler}, \citenamefont {Bimali}, \citenamefont {Yu},\ and\ \citenamefont {Fridley}}]{chalise_2014}%
  \BibitemOpen
  \bibfield  {author} {\bibinfo {author} {\bibfnamefont {P.}~\bibnamefont {Chalise}}, \bibinfo {author} {\bibfnamefont {D.~C.}\ \bibnamefont {Koestler}}, \bibinfo {author} {\bibfnamefont {M.}~\bibnamefont {Bimali}}, \bibinfo {author} {\bibfnamefont {Q.}~\bibnamefont {Yu}}, \ and\ \bibinfo {author} {\bibfnamefont {B.~L.}\ \bibnamefont {Fridley}},\ }\href {\doibase 10.3978/j.issn.2218-676X.2014.06.03} {\bibfield  {journal} {\bibinfo  {journal} {Transl Cancer Res}\ }\textbf {\bibinfo {volume} {1}} (\bibinfo {year} {2014}),\ 10.3978/j.issn.2218-676X.2014.06.03}\BibitemShut {NoStop}%
\end{thebibliography}%

\appendix
\section{Classical machine learning models}\label{sec:mlpipeline}
Multiple regression models are trained on both classical and quantum-embedded training data. The performance of these models is evaluated using metrics such as the Mean Squared Error (MSE). The program compares the performance of classical raw features, Gaussian Process Regressor with RBF kernel embedded data, and data embedded with Quantum Reservoir Computing (QRC) to assess the potential performance benefits of quantum and classical data embedding methods.

Although in the final analysis we used only random forest regressor and Gaussian RBF kernel, we tested many other models to see if they performed better on our data. The following regression algorithms were included in this study:
\begin{enumerate}
    \item Decision Tree Regressor: A non-parametric supervised learning method used for regression. This model predicts the value of a target variable by learning simple decision rules inferred from the data features~\cite{decisiontree}.

    \item Bagging Regressor: An ensemble meta-estimator that fits base regressors each on random subsets of the original dataset and then aggregates their individual predictions to form a final prediction. This is typically used to reduce variance for algorithms like decision trees~\cite{bagreg}.

    \item Gradient Boosting Regressor: An ensemble technique that builds models sequentially, each new model attempting to correct errors made by the previous ones. Models are built using the gradient descent algorithm to minimize the loss~\cite{gradboost}.

    \item Random Forest Regressor: An ensemble of decision trees, typically trained via the bagging method. It is used for regression by averaging the predictions of each component tree~\cite{randomforestreg}.

    \item AdaBoost Regressor: An adaptive boosting algorithm that adjusts the weights of incorrectly predicted instances so that subsequent classifiers focus more on difficult cases~\cite{adaboost}.

    \item Gaussian Process Regressor (GPR) with a Radial Basis Function (RBF) kernel: Effective for capturing smooth, non-linear relationships in the data. The RBF kernel with an added level of noise allows the model to provide both predictions and uncertainty estimates~\cite{gaussianprocessregressor}.

    \item Linear Regression: A linear approach to modeling the relationship between a dependent variable and one or more independent variables~\cite{linreg}.

    \item Support Vector Regression (SVR): An extension of the Support Vector Machine (SVM) that supports linear and non-linear regression~\cite{svr}.

    \item K-Neighbors Regressor: A type of instance-based learning or non-generalizing learning that does not attempt to construct a general internal model but simply stores instances of the data~\cite{kneighbor}.

    \item Extra Tree Regressor: Implements a meta estimator that fits a number of randomized decision trees on various sub-samples of the dataset and uses averaging to improve the predictive accuracy and control over-fitting~\cite{exttree}.

    \item SGD Regressor: A linear model fitted by minimizing a regularized empirical loss with SGD, the gradient of the loss is estimated each sample at a time, and the model is updated along the way with a decreasing learning rate~\cite{sgdr}.
    
    \item Neural Network: We employed a four-layer neural network for regression to approximate the depth of the QRC regression model. The model sequentially processes input data, extracting and consolidating features through the hidden layers, and employs a rectified linear unit activation function to predict continuous target values in the output layer~\cite{Schmidhuber_2015}.
\end{enumerate}

\begin{figure*}[!htb]    
    \centering
    \includegraphics[width=0.95\linewidth]{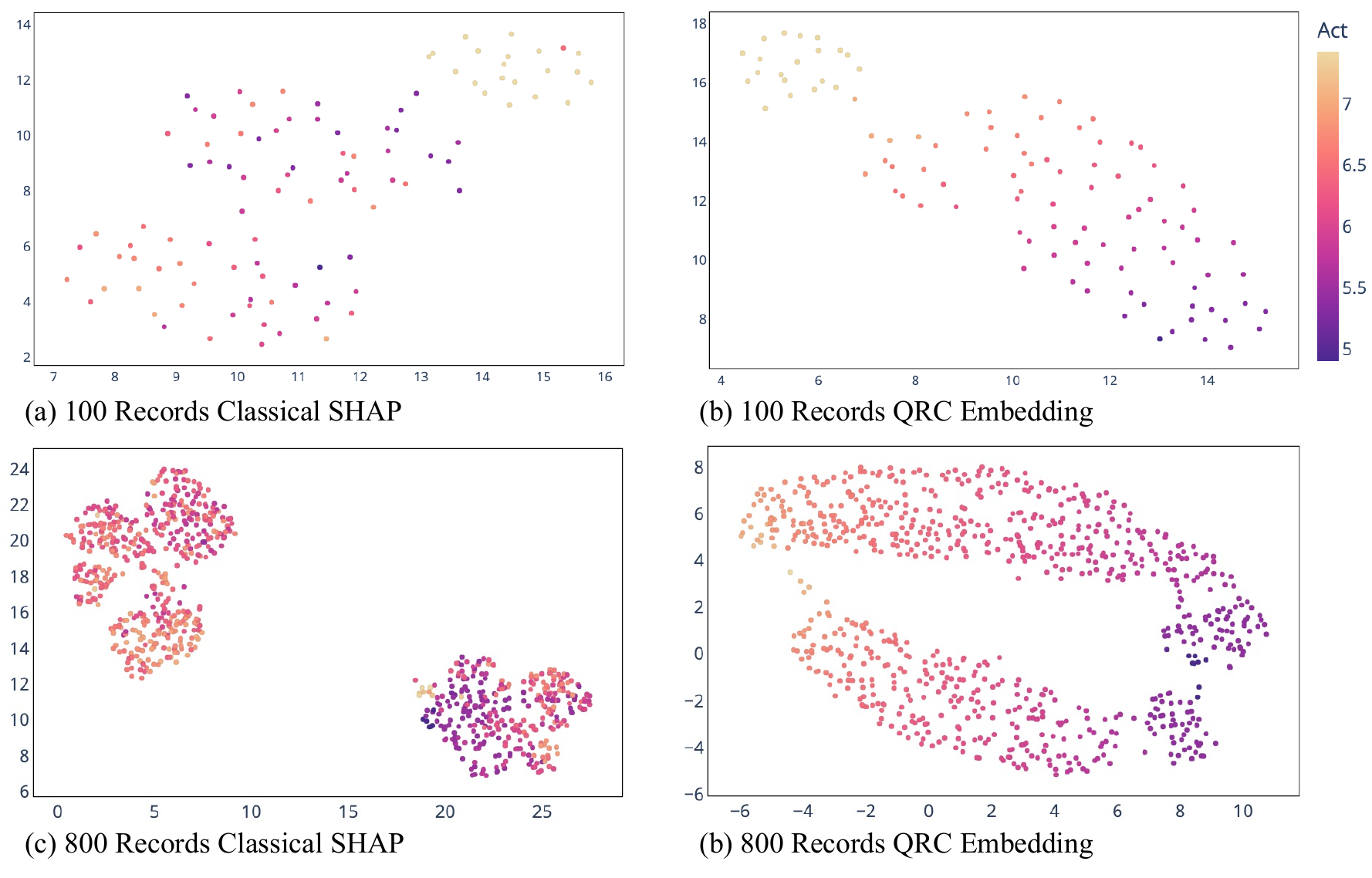}
    \caption{\justifying \textbf{UMAP 2D projections for 100 and 800 record samples of MMACD dataset 4.}}
    \label{fig:umap2d-100800graphs}
\end{figure*}

\section{Details of the modeling pipeline}

\subsection{Data Preparation Steps\label{sec:dataprep_appendix}}

\textbf{Loading and Cleaning Data}: The program begins by loading the dataset and performing necessary cleaning operations. This includes removing any unnamed columns that may have been introduced during data export or import processes.

\textbf{Exploratory Data Analysis:} The initial exploratory data analysis steps of the program involve preparing the dataset and selecting relevant features for the analysis. The dataset used in this study is derived from the MMACD challenge, which includes various molecular descriptors and activity labels. We examine the univariate statistics for individual columns in boxplots, scatterplots check for missing values, outlier values, as well as skew and kurtosis of the data distribution. 

\textbf{Data Standardization}: Next the dataset undergoes standardization to ensure that all features are on a comparable scale, which is essential for the performance of many machine learning algorithms. The process begins by selecting a subset of the dataset, retaining only the columns deemed consistent and relevant for the analysis, while excluding any non-essential columns such as identifiers or target variables. This selected subset is then standardized so that each feature has a mean of zero and a standard deviation of one using scikit-learn's standardscalar function~\cite{scikit-learn}. This normalization step is crucial for mitigating the effects of differing scales among features and ensuring that each feature contributes equally to the model's performance. We then use scikit-learn's train-test-split function to divide the data into training and test data sets of the desired size~\cite{scikit-learn}.

\subsection{Initial Candidate Model}\label{sec:cand_mods}

In the initial candidate model evaluation phase, several key metrics are initialized to systematically assess the performance of the machine learning models. The training dataset is based on raw classical features that are run against all models in the machine learning pipeline (See Fig.~\ref{sec:mlpipeline}). The model's performance metrics are compared against one another, including MSE. We select the model with the best performance as our initial candidate model, which we then use as the basis for the SHAP feature selection based on feature importance.

\subsection{SHAP Feature Selection}\label{shappapp}

Our pipeline utilizes SHAP (SHapley Additive exPlanations) values to identify important features. The selected features are then used to create a subset of the original dataset, ensuring that only the most relevant descriptors are included in the subsequent analysis.  As seen in Fig.~\ref{fig:QRC-workflow}, after our initial classical models were created, we applied SHAP to the data and determined the most important features for our champion model. The original MMACD  dataset 4 has 4,308 columns which would correspond to 4,308 qubits when encoded with the QRC. Due to the limited computing resources, we were able to perform the QRC simulation for 18 SHAP features out of 4,308, corresponding to 18 qubits. The SHAP kernel method fits a specified model, in our case, the Random Forest regressor, which performed the best on the entire data set, to our data~\cite{10.5555/3295222.3295230}. The SHAP kernel method explains the model's predictions by computing SHAP values for the training data, which quantify the contribution of each feature to the model's predictions~\cite{10.5555/3295222.3295230}. Following the computation of SHAP values, we identify the most important features by averaging the absolute SHAP values and ordering them by relative importance. We then select the top features to retain for further analysis.

\subsection{Clustering Subsample Algorithm}
Next, in our research modeling and visualization workflow (See Fig.~\ref{fig:QRC-workflow}), we create clustering-based subsamples. Our sampling methodology helps create subsamples that can be efficiently processed on our quantum simulator while at the same time being representative of the entire dataset without the risk of the same record being in multiple subsamples~\cite{chalise_2014}. To create valid subsamples of the MMACD dataset 4, we created a methodology for subsampling that clusters data in order to improve the representativeness of the subsample to the total population 
~\cite{chalise_2014}. We remove the target variable, then we apply the KMeans clustering algorithm to the remaining numeric data, predicting cluster assignments for each data point. This clustering step helps in grouping similar data points together. Next, we calculate the sample size per cluster and sample data points from each cluster with no replacement so each record is included in only one subsample~\cite{chalise_2014}. Our sampling methodology separates the features and target values from the sampled subset and splits this subset into training and testing sets, stratified by the cluster predictions. The optional modeling step is creating the QRC embeddings, as described in more detail in the section \ref{sec:QRC_methods}. After that, we return to the modeling pipeline to put the QRC embedded data, Gaussian Process Regressor with RBF kernel, and raw classical features as training and test data into the models as input data.

 \section{Supplementary UMAP Results}\label{sec:umapres100800}
 
The patterns in the 2D UMAP representation of the data, as presented in Fig.~\ref{fig:umap2d-100800graphs} largely mirror the patterns shown in the main text of the article for the 200 record sample, except that the pattern is less manifest in the 800 record samples. For 100 records, as seen in Fig.~\ref{fig:umap_200recs}, there is an area of high ACT concentration for the classical data, while for the QRC two-body observable embedded data there is a clearer pattern in the data for all activity values. The same can be seen in the 800 record sample, with classical data having a tendency towards high values clustering to the lower values of the X-axis but otherwise, no other clear pattern. For the 800 record QRC two-body observable embeddings (Fig.~\ref{fig:umap2d-100800graphs}), there is a clear pattern of the highest values being on the left and a decrease in ACT values towards the higher values of the X-axis, albeit with a more complex data cluster topology. The 100 and 800 record samples reinforce the observation of the clearer structure of the QRC embedded features versus the raw classical features.

\end{document}